\def\dint{\mathop{\displaystyle \int}}

\newcommand\strt[1]{\rule[-#1pt]{0pt}{#1pt}}

\newcommand\eq{=&&\hspace{-18pt}}

\documentclass[12pt,a4paper]{article}
\usepackage{amsmath}
\usepackage{graphicx}
\usepackage[usenames,dvipsnames]{xcolor}
\usepackage{mathpazo} 
\usepackage{hyperref}
\usepackage{multimedia}
\usepackage{comment}
\usepackage[small,nohug,heads=vee]{diagrams}
\diagramstyle[labelstyle=\scriptstyle]
\definecolor{light-gray}{gray}{0.80}

\usepackage{fancyhdr}
\renewcommand{\maketitle}{
    \begin{center}
      \Large
        {\bf Speeds of light and mass stability in Stueckelberg-Horwitz-Piron electrodynamics}
        \vskip .3 true cm
      \small
        Martin Land \\
        \vskip .3 true cm
        Department of Computer Science \\
        Hadassah College \\
        37 HaNevi'im Street, Jerusalem \\
email: martin@hadassah.ac.il
      \end{center}
      \vskip .5 true cm
}
\input{tcilatex}

\begin{document}
\title{}
\author{}
\maketitle
%


%
\begin{abstract}
It is well-known that the 5D gauge structure of Stueckelberg-Horwitz-Piron (SHP)
electrodynamics permits the exchange of mass between particles and the fields
induced by their motion, even at the classical level.  This phenomenon presents
two closely related problems: (1) What accounts for the stability of the
measured masses of the known particles?  (2) Under what circumstances can real
particles evolve sufficiently off-shell to account for mass changing phenomena
such as flavor-changing neutrino interactions and low energy nuclear reactions?
To approach these questions, we introduce a constant $c_5$ associated with the
invariant time $\tau$, in analogy with the constant $c$ that associates a unit
of length with intervals of time $t$ in standard relativity.  It follows that
electromagnetic mass exchange can be a small effect, in proportion to $c_5 / c$.
We show that this structure permits a classical self-interaction that tends to
restore on-shell propagation.  Finally we propose a model in which a particle
evolving through a complex charged environment can acquire a significant mass shift for a
short time.  

\end{abstract}

\baselineskip7mm \parindent=0cm \parskip=10pt

\section{Introduction}

In a formal approach to special relativity that takes Minkowski geometry as its
starting point, the constant
$c$ is introduced as a means of measuring time in units of spatial distance,
and the notion of a speed of light emerges from the role of $c$ in wave
equations for U(1) gauge fields.  Using natural units ($\hbar = c = 1$) in the
development of SHP electrodynamics \cite{Stueckelberg} - \cite{H-11}, no explicit constant was assigned to the
invariant time $\tau$ and so the constant $c$ was implicitly assumed to play
the same role for $\tau$ that it plays for the coordinate time $t$.  In 
Section 2 we associate a new constant $c_5$ with the invariant time $\tau$, identify
the expressions in which it must appear and study its role in the classical
electromagnetic theory.  Unlike standard 4D special relativity, in which the
nonrelativistic limit can be recovered by taking $c \rightarrow \infty$, we find
that 5D SHP goes over to an equilibrium state of Maxwell theory in the limit
$c_5 \rightarrow 0$.  Thus, the dimensionless ratio $c_5/c$ parameterizes the
deviation of SHP from standard electrodynamics, in particular the coupling of
the events that dynamically trace out particle worldlines to the mass changing
fields.  Put another way, equilibrium Maxwell theory can be understood as 
the pre-Maxwell fields becoming independent of $c_5 \tau$ as $c_5 \rightarrow 0$. 

In Section 3 we construct a model for the self-interaction involving an
event and the causally retarded field produced by its motion.  In numerical
solutions \cite{jigal} it was found that under interactions of this
type, the particle mass may asymptotically approach its on-shell value.  Here we
calculate the classical Lorentz force produced by the self-interaction in the
particle rest frame --- it is seen to produce a damping force tending to return
an off-shell event to on-shell evolution, and vanishing for on-shell evolution.
In Section 4 we propose a simple mechanism by which a particle evolving through
a complex plasma may acquire a significant mass shift for a short time.  

\section{Overview of Stueckelberg-Horwitz-Piron (SHP) electrodynamics}
\subsection{Gauge theory}

This section covers familiar territory, reformulated to make each physical
constant explicit.  The generalized Stueckelberg-Schrodinger equation
\begin{equation}
(i\hbar \partial _{\tau }+\frac{e_{0}}{c}\phi )\ \psi (x,\tau )=\frac{1}{2M}%
(p^{\mu }-\frac{e_{0}}{c}a^{\mu })(p_{\mu }-\frac{e_{0}}{c}a_{\mu })\ \psi
(x,\tau )
\label{schr}
\end{equation}
describes the interaction of an event characterized by the wavefunction
$\psi(x,\tau)$ with five gauge fields $a_\mu(x,\tau)$ and $\phi(x,\tau)$.  Equation
(\ref{schr}) is invariant under the local gauge transformations
\begin{equation}
\begin{array}{lrcl}
&\psi (x,\tau )&\rightarrow&
\exp \left[ \dfrac{ie_{0}}{\hbar c}\Lambda (x,\tau )\right] \, \psi (x,\tau
)\vspace{8pt} \\
\text{Vector potential }\quad&a_{\mu }(x,\tau )&\rightarrow& a_{\mu }(x,\tau
)+\partial _{\mu }\Lambda (x,\tau )\vspace{8pt} \\
\text{Scalar potential }\quad&\phi(x,\tau )&\rightarrow& \phi(x,\tau )+\partial _{\tau }\Lambda (x,\tau ) \\
\label{l-gauge}
\end{array}
\end{equation}
whose $\tau$-dependence is the essential departure from Stueckelberg's work, and
determines the structure of the resulting theory \cite{saad}.  The corresponding
global gauge invariance leads to the conserved Noether current  
\begin{equation}
\partial _{\mu }j^{\mu }+\partial _{\tau }\rho = 0
\label{cons}
\end{equation}
where
\begin{equation}
j^{\mu }=-\dfrac{i\hbar}{2M}\left\{\psi ^*\left(\partial ^{\mu
}-\dfrac{ie_0}{c}a^{\mu }\right)\psi -\psi \left(\partial ^{\mu }+\dfrac{ie_0}{c}a^{\mu
}\right)\psi ^*\right\}\qquad \rho =\left\vert\psi (x,\tau )\right\vert^{2} \ \ .
\end{equation}
In analogy to the notation $x^0 = ct$ we adopt the formal designations
\begin{equation}
x^5 = c_5 \tau \qquad \qquad   \partial_5 = \dfrac{1}{c_5} \partial_\tau \qquad  
\qquad j^5 = c_5 \rho \qquad\qquad  a_5 =  \dfrac{1}{c_5} \phi
\end{equation}
and the index convention 
\begin{equation}
\lambda ,\mu ,\nu =0,1,2,3 \qquad \qquad \alpha ,\beta ,\gamma =0,1,2,3,5
\end{equation}
so that the gauge and current conditions (\ref{l-gauge}) and (\ref{cons}) can be written
\begin{equation}
a_{\alpha }\rightarrow a_{\alpha }+\partial _{\alpha }\Lambda 
\qquad \qquad 
\partial _{\alpha }j^{\alpha } = 0 \ \ .
\end{equation}
It is convenient to choose the factor $g_{55}$ and $g^{55} = 1/g_{55}$ to apply when raising and lowering the 5-index, so that
\begin{equation}
%
\partial _{\alpha }j^{\alpha } =
g^{\mu\nu} \partial _{\mu }j_{\mu } + g^{55} \partial _{5 }j_{5 } \ \ . 
\end{equation}

\subsection{Classical event dynamics}

The classical mechanics of a relativistic event is found by rewriting the 
Stueckelberg-Schrodinger equation in the form
\begin{equation}
i\hbar \partial _{\tau }\psi (x,\tau ) =\left[ \frac{1}{2M}(p^{\mu }-\frac{%
e_{0}}{c}a^{\mu })(p_{\mu }-\frac{e_{0}}{c}a_{\mu })-\frac{e_{0}}{c}\phi %
\right] \psi (x,\tau ) 
\end{equation}
and transforming the classical Hamiltonian to Lagrangian form as
\begin{equation}
L=\dot{x}^{\mu }p_{\mu }-K=\dfrac{1}{2}M\dot{x}^{\mu }\dot{x}_{\mu }+\dfrac{e_0}{c}
\dot{x}^{\alpha }a_{\alpha }
\end{equation}
where
\begin{equation}
\dot{x}^\mu = \dfrac{dx^\mu}{d\tau} \qquad \dot{x}^5 = \dfrac{dx^5}{d\tau}
\equiv c_5
\ \ .
\end{equation}
The Euler-Lagrange equations
\begin{equation}
\dfrac{d}{d\tau }\dfrac{\partial L}{\partial \dot{x}_{\mu }}
-\dfrac{\partial L}{\partial x_{\mu }}=0
\end{equation}
are
\begin{equation}
\frac{d}{d\tau }\left( M~\dot{x}^{\mu }+\frac{e_{0}}{c}a^{\mu }\right)
-\partial ^{\mu }\left( \frac{e_{0}}{c}\dot{x}^{a}a_{a}\right) = 0
\end{equation}
leading to the Lorentz force 
\begin{eqnarray}
M\ddot{x}^{\mu } \eq \dfrac{e_0}{c} \big[ \dot{x}^{\alpha} \partial^\mu a_\alpha -
\dot{x}^\alpha \partial_\alpha a^{\mu } \big]
= \dfrac{e_0}{c} f^\mu_{\; \;  \alpha} (x,\tau )\dot{x}^\alpha \strt{12}
\notag \\
\eq \dfrac{e_0}{c} f^\mu_{\; \;  \nu} (x,\tau )\dot{x}^\nu
+ \dfrac{e_0}{c} f^\mu_{\; \;  5} (x,\tau )\dot{x}^5 \strt{12}
\notag \\
\eq  \dfrac{e_0}{c} f^\mu_{\; \;  \nu} (x,\tau )\dot{x}^\nu
- g_{55} \dfrac{e_0c_5}{c} f^{5\mu} (x,\tau ) 
\label{lorentz}
\end{eqnarray}
where
\begin{equation}
f^\mu_{\; \;  \alpha} = \partial^\mu a_\alpha - \partial_\alpha a^\mu \ \ .
\end{equation}
Because the four components of $\dot x^\mu$ are independent, the event
evolution may be off-shell.  In this context, on-shell evolution obeys the
mass-shell constraint $\dot x^2 = -c^2$ of standard relativity.  In SHP
electrodynamics
\begin{equation}
\dot x^2=\left( c\frac{dt}{d\tau },\frac{d\mathbf{x}}{d\tau }\right) ^{2}=c^{2}
\dot{t}^{2}\left( 1,\frac{1}{c}\left(\frac{d\mathbf{x}}{d\tau
}\right)\left(\frac{dt}{d\tau }\right)^{-1}
\right) ^{2}=c^{2}\dot{t}^{2}\left( 1,\frac{1}{c}\frac{d\mathbf{x}}{dt}
\right) ^{2}=-c^{2}\dot{t}^{2}\left( 1-\frac{\mathbf{v}^{2}}{c^{2}}\right)
%
\end{equation}
so that an event evolves on-shell when
\begin{equation}
\left\vert\frac{dt}{d\tau } \right\vert = \left\vert\dot{t} \right\vert = \frac{1}{\sqrt{1-\frac{\mathbf{v}^{2}}{c^{2}}}}
\end{equation}
and is said to be off-shell when $\left\vert\dot{t} \right\vert$ takes any other value.
In the SHP formalism, particles may exchange mass with fields through
\begin{equation}
\frac{d}{d\tau }(-\tfrac{1}{2}M\dot{x}^{2})=-M\dot{x}^{\mu }\ddot{x}_{\mu
}=-\dfrac{e_0}{c}\; \dot{x}^{\mu }(c_5 f_{\mu 5}+f_{\mu \nu }\dot{x}^{\nu })=
\dfrac{e_0 c_5}{c}\; \dot{x }^{\mu }f_{5\mu }= g_{55} \dfrac{e_0 c_5}{c}\; f^{5\mu }\dot{x }_{\mu}
\label{mass-shift}
\end{equation}
and the mass shell is demoted from the status of constraint to
that of conservation law for interactions in which $\dot{x }^{\mu }f_{5\mu } =
0$ (which usually entails $f_{5\mu }=0$).  However, if the scale of the fields
$f_{5\mu }$ is small compared to the Maxwell fields $f_{\mu\nu }$ then the
exchange of mass will be correspondingly small.  It would be convenient to find
that
\begin{equation}
c_5 \rightarrow 0 \quad \Rightarrow \quad f_{5\mu } \rightarrow 0
\end{equation}
so that $c_5$ can be understood as the scale of dynamic evolution in the
microscopic system, approaching an equilibrium equivalent to standard Maxwell
theory as $c_5 \rightarrow 0 $.  With this expectation in mind we examine the
fields produced by the motions of charged events. 

\subsection{Electromagnetic action}

To write an
electromagnetic action requires the choice of a kinetic term for the gauge
field; this term must be both gauge and O(3,1) invariant.  We write
\begin{equation}
S_\text{em} = \int d^{4}xd\tau \left\{\dfrac{e_0}{c}j^{\alpha }(x,\tau )a_{\alpha }(x,\tau )-
\int ds\, \frac{\lambda}{4c} \left[f^{\alpha \beta }(x,\tau )\Phi (\tau -s)f_{\alpha \beta
}\left( x,s\right) \right] \right\}
\label{action}
\end{equation}
where the five components of the local event current 
\begin{equation}
j^\alpha (x,\tau) =  c\dot{X}^\alpha(\tau) \delta^4\left(x-X(\tau) \strt{3} \right)
\label{pm_curr}
\end{equation}
have support at the spacetime location $X^\mu (\tau)$ of the event, and we again write $\dot
X^5 = c_5$. 
The $\tau$-integral of (\ref{pm_curr}) along the worldline concatenates the instantaneous events
into the Maxwell particle current in the usual form.  The field interaction kernel is
defined as  
\begin{equation}
\Phi (\tau ) = \delta \left( \tau \right) -(\alpha\lambda) ^{2}\delta ^{\prime \prime
}\left( \tau \right) = \int \frac{d\kappa}{2\pi} \,\left[ 1+\left( \alpha\lambda
\kappa \right) ^{2}\right] \,e^{-i\kappa \tau } 
%
%
\label{kernel}
\end{equation}
where
\begin{equation}
\alpha =\frac{1}{2}\left[ 1+\left( \frac{c_5}{c}\right) ^{2}\right]
\label{alpha}
\end{equation}
is chosen so that the low energy Coulomb force agrees with the standard
expression.  The inverse function of the interaction kernel
\begin{equation}
\varphi (\tau ) = \Phi^{-1}  (\tau )= \int \frac{d\kappa}{2\pi}
\,\frac{e^{-i\kappa \tau }}{
1+\left(\alpha  \lambda \kappa \right) ^{2}}
=\frac{1}{2\alpha \lambda }e^{-\vert\tau \vert/\alpha \lambda }
\label{inv}
\end{equation}
%
satisfies
\begin{equation}
\int d\tau ~\varphi \left( \tau \right)  =1
\end{equation}
and appears in the field equations as a smoothing of the particle current with
respect to the sharp location of each individual event.
%

Expanding the expression
\begin{equation}
f^{\alpha\beta}f_{\alpha\beta} = f^{\mu\nu}f_{\mu\nu} + 2 f^{5\mu}f_{5\mu} 
= f^{\mu\nu}f_{\mu\nu} + 2 g^{55} f_{5}^{~\mu}f_{5\mu}
\end{equation}
we may side-step the interpretation of $g^{55}$ as an element in a 5D metric and rather
see its role as equivalent to the choice of sign for
the vector contribution $f_{5}^{~\mu}f_{5\mu}$ to the field energy. 
Using (\ref{kernel}) and integrating by parts, the action takes the form
\begin{equation}
S = \int d\tau \; \frac{1}{2}M\dot{x}^{\mu }\dot{x}_{\mu }
+ \int d^4x d\tau \left\{ \dfrac{e_0}{c} \, a_\alpha  j^\alpha 
-\frac{\lambda }{4c}f_{\alpha \beta }f^{\alpha \beta }
-\frac{\alpha^2\lambda ^{3}}{4c}\left(\partial _{\tau }f^{\alpha \beta
}\right)\left(\partial _{\tau }f_{\alpha \beta }\right)
\right\}
\end{equation}
in which the gauge and O(3,1) invariance are manifest, but the $\tau$ derivatives
in the last term
explicitly break any formal 5D symmetry of the terms $f_{\alpha \beta }f^{\alpha \beta }$.  

Varying the action in the form (\ref{action}) with respect to the fields,
and using (\ref{inv}) to remove the kernel $\Phi$, leads to the field equations
\begin{eqnarray}
&&\partial _{\beta }f^{\alpha \beta }\left( x,\tau \right)
= \frac{e_{0}}{\lambda c}\int ds~\varphi \left( \tau -s\right)
j^{\alpha }\left( x,s\right)
= \dfrac{e}{c} \, j_\varphi^{\alpha } \left( x,\tau \right) \strt{22}
\label{gauss}\\
&&\partial _{\alpha }f_{\beta \gamma } + 
\partial _{\gamma }f_{\alpha \beta } + 
\partial _{\beta }f_{\gamma \alpha }  = 0 
\label{pm-h}
\end{eqnarray}
which are formally similar to 5D Maxwell equations with $e = e_0 / \lambda$.
The source of the field in (\ref{gauss}) is 
\begin{equation}
j_\varphi^{\alpha } \left( x,\tau \right) = \int ds \ \varphi(\tau -s) j^{\alpha } \left( x,\tau \right)
= c\int ds \ \varphi(\tau -s) \dot{X}^\alpha(s) \delta^4\left(x-X(s) \strt{3} \right)
\end{equation}
formed by smoothing the support of the
instantaneous current $j^{\alpha } \left( x,\tau \right) $ defined in
(\ref{pm_curr}) by convolution with the inverse kernel function
$\varphi(\tau)$.  For $\lambda$ very small, $\varphi$ becomes a delta function
which narrows the source to a small neighborhood around the event inducing the
current.  For $\lambda$ very large, the convolution becomes a concatenation of
the current along the worldline, equivalent to the Maxwell current.  The
parameter $\lambda$ thus plays the role of a correlation length,
characterizing the range of the electromagnetic interaction.   

The field equations (\ref{gauss}) and (\ref{pm-h})
are called pre-Maxwell equations, and together with the Lorentz force
(\ref{lorentz}) describe a microscopic event dynamics for which Maxwell theory
can be understood as an equilibrium limit.  
The connection with Maxwell theory is found
by integration over $\tau$ which concatenates the events along the worldline.  With
equilibrium boundary conditions
\begin{equation}
\rho_\varphi (x,\tau) \underset{\tau \rightarrow \pm \infty }{\mbox{\
}\xrightarrow{\hspace*{1.5cm}}\mbox{\ }} 0 \qquad \qquad 
f^{5\mu} (x,\tau) \underset{\tau \rightarrow \pm \infty }{\mbox{\
}\xrightarrow{\hspace*{1.5cm}}\mbox{\ }} 0
\end{equation}
we find
\begin{equation}
\left. 
\begin{array}{c}
\partial _{\beta }f^{\alpha \beta }\left( x,\tau \right)
=\dfrac{e}{c} j_{\varphi}^{\alpha }\left( x,\tau \right) \\ 
\\ 
\partial _{\lbrack \alpha }f_{\beta \gamma ]}=0 \\
\\ 
\partial _{\alpha }j^{\alpha } = 0
\end{array}
\right\}
\underset{\int d\tau }{\mbox{\quad}\xrightarrow{\hspace*{1cm}} \mbox{\quad}}\left\{ 
\begin{array}{c}
\partial _{\nu }F^{\mu \nu }\left( x\right) =\dfrac{e}{c}J^{\mu }\left( x\right) \\ 
\\ 
\partial _{\lbrack \mu }F_{\nu \rho ]}=0 \\
\\ 
\partial _{\mu } J^{\mu}(x) = 0
\end{array}
\right.
\end{equation}
where
\begin{equation}
%
A^{\mu }(x)=\int d\tau \;a^{\mu }(x,\tau )
\qquad F^{\mu \nu }(x)=\int d\tau \;f^{\mu \nu }(x,\tau )
\qquad J^{\mu}(x) = \int d\tau \; j^{\mu }(x,\tau ) \ \ .
\end{equation}
Since $e_0a^\mu$ must have the dimensions of $eA^\mu$, it follows that 
$e_{0}$ and $\lambda$ have the dimension of time and $e=e_{0}/\lambda$ is dimensionless.

Rewriting the field equations in vector and scalar components, they take the form
\vspace{4pt}
\begin{equation}
\begin{array}{lcl}
\partial _{\nu }\;f^{\mu \nu }- \dfrac{1}{c_5} \partial _{\tau }\;f^{5\mu
}=\dfrac{e}{c} \; j^{\mu }_\varphi 
& \mbox{\qquad} &\partial _{\mu }\;f^{5\mu }=\dfrac{e}{c}\;  j^5_\varphi
= \dfrac{c_5}{c} \; e \rho_\varphi
\vspace{8pt}\\ 
\partial _{\mu }f_{\nu \rho }+\partial _{\nu }f_{\rho \mu }+\partial _{\rho }f_{\mu \nu }=0 
& &\partial _{\nu }f_{5\mu }-\partial _{\mu }f_{5\nu }
+ \dfrac{1}{c_5} \partial _{\tau }f_{\mu \nu }=0 \\
\end{array}
\label{premax}
\vspace{4pt}
\end{equation}
which may be compared with the 3-vector form of Maxwell equations
\begin{equation}
\vspace{4pt}
\begin{array}{lcl}
\nabla \times \mathbf{B}-\dfrac{1}{c} \partial _{t}\mathbf{E}=\dfrac{e}{c} \ \mathbf{J} 
& \mbox{\qquad} \mbox{\qquad} & \nabla \cdot \mathbf{E}=\dfrac{e}{c} \ J^{0} \vspace{8pt}\\ 
\nabla \cdot \mathbf{B}=0
& &\nabla \times \mathbf{E}+\dfrac{1}{c} \partial _{t}\mathbf{B}=0 \\
\end{array}
%
\end{equation}
with $f_{5\mu }$ playing the role of the vector electric field and $f^{\mu \nu
}$ playing the role of the magnetic field.  We notice that $c_5$ appears three
times in the pre-Maxwell equations (\ref{premax}), twice in the form $\frac{1}{c_5} \partial
_{\tau }$ and once multiplying the event density $\rho_\varphi$.  To make sense
of the derivative terms, we first recall that the homogeneous pre-Maxwell
equations are automatically satisfied for fields derived from potentials --- in
this case the fields $f_{5\mu}$ contain terms with $\partial_5 a_\mu =
\frac{1}{c_5} \partial_\tau a_\mu$ that cancel the explicit $\tau$-derivative of
$f_{\mu\nu}$.  From the second homogeneous equation it follows that
\begin{equation}
0=c_5 \left( \partial _{\nu }f_{5\mu }-\partial _{\mu }f_{5\nu } \right) +
\partial _{\tau }f_{\mu \nu }\underset{ c_5 \rightarrow 0
}{\mbox{\ }\xrightarrow{\hspace*{1cm}} \mbox{\ }} \partial _{\tau }f_{\mu
\nu }
\end{equation}
so that the Maxwell field strength becomes $\tau$-static.
The $\tau$-derivative term in the first inhomogeneous pre-Maxwell
equation remains finite as long as $f^{5\mu}$ is proportional to $c_5$ and we
will see that this is generally the case for fields derived from potentials of
the Li\'{e}nard-Wiechert type.  Under the boundary conditions associated with
concatenation, the event density $\rho_\varphi$ and $f^{5\mu}$ both vanish in
equilibrium.  Under the slightly weaker assumption that the divergenceless free
field $f^{5\mu}$ is $\tau$-independent, it decouples from the Maxwell field, so
that $f^{\mu\nu}$ and $j^\mu$ satisfy the standard Maxwell equations.
It is sometimes notationally convenient to further expand the field into 3-vector components as
\begin{equation}
(\mathbf{e})^i = f^{0i} \qquad (\mathbf{h})_{in} = \epsilon_{ijk} f^{jk} \qquad
(\mathbf{f^5})^i = f^{5i} \ \ .
\end{equation}

\subsection{Wave equations and induced fields}

The pre-Maxwell equations, in Lorenz gauge, lead to the wave equation 
\begin{equation}
\partial _{\beta }\partial ^{\beta }a^{\alpha }=(\partial _{\mu }\partial
^{\mu }+\partial _{\tau }\partial ^{\tau })a^{\alpha }=(\partial _{\mu
}\partial ^{\mu } +  \frac{g_{55}}{c_5^{2}} \; \partial _{\tau }^{2})a^{\alpha }=-
\frac{e}{c}\ j_{\varphi }^{\alpha }\left( x,\tau \right)
\end{equation}
whose solutions may respect 5D symmetries broken by the O(3,1) symmetry of the event
dynamics.  A Green's function solution to
\begin{equation}
(\partial_{\mu}\partial^{\mu} + \frac{g_{55}}{c_5^{2}} \; \partial _{\tau }^{2})G(x,\tau)
=- \delta^4\left( x \right) \delta \left( \tau \right)
\end{equation}
can be used to obtain potentials of the form
\begin{eqnarray}
a^{\alpha }\left( x,\tau \right) \eq -\frac{e}{c} \int d^{4}x^{\prime }d\tau
^{\prime } \ G\left( x-x^{\prime }, \tau -\tau ^{\prime }\right) j_{\varphi
}^{\alpha }\left( x^{\prime },\tau^{\prime } \right) \notag
\\
\notag
\eq -e \int d^{4}x^{\prime}d\tau^{\prime }ds \ G\left( x-x^{\prime }, \tau -\tau ^{\prime }\right) 
\varphi(\tau^{\prime} -s) \dot{X}^\alpha(s) \delta^4\left(x^{\prime}-X(s) \strt{3} \right)\\
\notag
\eq -e \int ds \left[ \int d\tau^{\prime }\ G\left( x-X(s), \tau -\tau ^{\prime
}\right) \varphi(\tau^{\prime} -s) \right]
\dot{X}^\alpha(s) \\
\eq -e \int ds \ G_{\varphi} \left( x-X(s), \tau -s\right) \ \dot{X}^\alpha(s) \ .
\end{eqnarray}
Since $\dot{X}^5(s) = c_5$, while the 4-vector $\dot{X}(s) = \dot{X}^0 (s)
(c, {\mathbf v}) $ with $\vert {\mathbf v}\vert < c$, we see that the fifth
potential $a^5(x,\tau)$ is in general scaled by $c_5/c$ with respect to $a^\mu(x,\tau)$.

The principal part Green's function was found \cite{green} using Schwinger's
method in the form
\begin{eqnarray}
G_{P}(x,\tau ) \eq -{\frac{1}{{4\pi }}}\delta (x^{2})\delta (\tau )-{\frac{c_5}{{
2\pi ^{2}}}}{\frac{\partial }{{\partial {x^{2}}}}}{\theta (-g_{55}g_{\alpha
\beta }x^{\alpha }x^{\beta })}{\frac{1}{\sqrt{-g_{55}g_{\alpha \beta
}x^{\alpha }x^{\beta }}}}
\label{greens} 
\\  \eq \  G_{Maxwell} + G_{Correlation}
\end{eqnarray}
which recovers the 4D Maxwell Green's function 
\begin{equation}
\int d\tau \ G_{Maxwell} = D(x) =  -{\frac{1}{{4\pi }}}\delta (x^{2})
\qquad 
\int d\tau \ G_{Correlation} = 0 
\end{equation}
under concatenation\footnote{The Green's function was derived taking $c = c_5 = 1$, so
that $x^5 = \tau$.  Working through the derivation and replacing $\tau$ with
$x^5 = c_5 \tau$ leads to a factor of $c_5$ multiplying the second term, so that
both terms have units of distance$^{-2} \; \times$ time$^{-1}$.}. The support of $G_{Correlation}$ is 
\begin{equation}
{-g_{55}g_{\alpha \beta }x^{\alpha }x^{\beta }=}\left\{ 
\begin{array}{lll}
-\left( x^{2}+c_{5}^{2}\tau ^{2}\right) =c^{2}t^{2}-\mathbf{x}%
^{2}-c_{5}^{2}\tau ^{2}>0 & , & {g_{55}=1} \strt{12}\\ 
\left( x^{2}-c_{5}^{2}\tau ^{2}\right) =\mathbf{x}^{2}-c^{2}t^{2}-c_{5}^{2}%
\tau ^{2}>0 & , & {g_{55}=-1}%
\end{array}%
\right.
\end{equation}
leading to causality properties discussed in \cite{green}.  In particular, we
see that for $g_{55} = 1$, the second term $G_{Correlation}$ has timelike support with respect to the
event trajectory, opening the possibility of a self-interaction of a type not
present in standard Maxwell theory.  In order to exploit this self-interaction,
we take $g_{55} = 1$ in the remaining sections of this paper.  

As required in Schwinger's method, we take special care in handling the
distribution functions and the order of integration.  Evaluating the derivative
in (\ref{greens}) we find 
\begin{eqnarray}
G_{correlation}\left( x,\tau \right) \eq -{\frac{c_5}{{2\pi ^{2}}}}{\frac{
\partial }{{\partial {x^{2}}}}}{\theta (-g_{55}g_{\alpha \beta }x^{\alpha
}x^{\beta })}{\frac{c_5}{\sqrt{-g_{55}g_{\alpha \beta }x^{\alpha }x^{\beta }}}}
\strt{18}
\notag \\
\eq
-{\frac{c_5}{{2\pi ^{2}}}}{\frac{\partial }{{\partial {x^{2}}}}\frac{{\theta (
{-x^{2}}-c_{5}^{2}\tau ^{2})}}{ \left( {-x^{2}}-c_{5}^{2}\tau ^{2}\right) ^{1/2} }}
\strt{18}
\notag \\
G_{correlation}\left( x,\tau \right) \eq -{\frac{c_5}{{2\pi ^{2}}}}\left( \frac{
1}{2}{\frac{{\theta ({-x^{2}}-c_{5}^{2}\tau ^{2})}}{\left( {-x^{2}}
-c_{5}^{2}\tau ^{2}\right) ^{3/2}}}-{\frac{{\delta }\left( {-x^{2}}
-c_{5}^{2}\tau ^{2}\right) }{\left( {-x^{2}}-c_{5}^{2}\tau ^{2}\right) ^{1/2}
}}\right) \ \ .
\end{eqnarray}
Although the second term appears highly singular, we will see that when
calculating potentials, singularities in the terms of $G_{correlation}$
cancel when the subtraction is performed before applying the limits of
integration. 

The `static' Coulomb potential in this framework is induced by an
isolated event moving uniformly along the $t$ axis.  Writing the event as
\begin{equation}
X\left( \tau \right) =\left(c \tau ,0,0,0\right)
\end{equation}
produces the currents
\begin{eqnarray}
j^0(x,\tau) \eq j^5 (x,\tau) =  c \delta(t-\tau) \, \delta^3({\mathbf x}) \qquad \qquad {\mathbf j}(x,\tau) = 0 \\
j_\varphi^0(x,\tau) \eq  j_\varphi^5 (x,\tau) =  c \varphi(t-\tau) \, \delta^3({\mathbf x}) \qquad \qquad {\mathbf j}_\varphi(x,\tau) = 0
\end{eqnarray}
and the Maxwell part of the Green's function induces 
\begin{equation}
a^{0} (x,\tau) =  a^{5}(x,\tau) 
={\frac{e}{{4\pi \vert \mathbf{x} \vert }}}\varphi
\left(\tau - \left( t - \frac{\vert \mathbf{x} \vert}{c} \right) \right)  \qquad \qquad \mathbf{a} = 0
\end{equation}
which recovers the standard Coulomb potential 
\begin{equation}
A^{0}(x)=\dint d\tau \, a^{0}\left( x,\tau \right) = {\dfrac{e}{{4\pi \vert
\mathbf{x} \vert}}}  \qquad \qquad \qquad \mathbf{A} = 0
\end{equation}
under concatenation.  In Appendix A we show that the contribution from $G_{Correlation}$
is smaller than the $G_{Maxwell}$ contribution by $c_5 / c$ and drops off as
$1/\left\vert \mathbf{x} \right\vert^2$.  A test event located at $x(\tau )=(c\tau ,\mathbf{x})$
will see the Yukawa-type potential
\begin{equation}
a^{0}(x,\tau )  =  a^{5}(x,\tau ) 
={\frac{e}{{4\pi \vert \mathbf{x} \vert }}}\frac{1}{2\alpha \lambda }
e^{-\left\vert \mathbf{x}\right\vert /\alpha \lambda c}
\label{yukawa}
\end{equation}
in which $1 / \lambda$ parameterizes the mass spectrum of the pre-Maxwell field.  If
$\lambda$ is small (so that $\varphi$ approaches a delta function and the current
narrows to around the event) the mass spectrum becomes wide.  If $\lambda$ is large,
the support of the current spreads along the worldline and the potential becomes
Coulomb-like. 
The field strength components are
\begin{equation}
f^{k0}(x,\tau )  = f^{k5}(x,\tau ) =\partial ^{k}{\frac{e}{{4\pi \vert\mathbf{x}\vert}}}\frac{1}{2\alpha \lambda }
e^{-\left\vert \mathbf{x}\right\vert /\alpha \lambda c} \qquad 
f^{ij}(x,\tau ) =0 \qquad f^{50} = 0
\end{equation}
where we used (\ref{inv}) for $\varphi\left(\tau\right)$.  The test event will 
experience the Coulomb force through (\ref{lorentz}) as 
\begin{equation}
M\ddot{x}^{k}=\dfrac{e_{0}}{c}f_{\;\;\nu }^{k}\dot{x}^{\nu }-g_{55}
\dfrac{e_{0}c_{5}}{c}f^{5k} = -\dfrac{e_{0}}{c}f^{k0}\left( \dot{x}
^{0}-g_{55}c_{5}\right)
\label{lor-coul}
\end{equation}
and since $\dot{x}(\tau )=(c,\mathbf{0})$ this becomes
\begin{equation}
M\mathbf{\ddot{x}} =-\frac{e_{0}e}{2\alpha \lambda }\left( 1-g_{55}\frac{c_{5}}{c}
\right) \nabla \left( \dfrac{e^{-\left\vert \mathbf{x}\right\vert /\alpha \lambda c}
}{{4\pi }\left\vert \mathbf{x}\right\vert }\right) 
=-e^{2}~\frac{1-g_{55}\frac{c_{5}}{c}}{1+\left( \frac{c_{5}}{c}\right) ^{2}
}\nabla \left( \dfrac{e^{-\left\vert \mathbf{x}\right\vert /\alpha \lambda c}}{{
4\pi }\left\vert \mathbf{x}\right\vert }\right) 
\label{lor-coul-2}
\end{equation}
where we used (\ref{alpha}) for $\alpha$.  This expression for the
Coulomb force would vanish for $c_5 = c$ and $g_{55} = 1$,  
and for this reason it was previously argued \cite{larry} that $g_{55} = 1$ 
(corresponding to a formal O(4,1) symmetry of the wave equation) is 
prohibited.  However, with $c_5 < c$ either signature for $g_{55}$ is permitted.
In (\ref{lor-coul}) $\dot x (\tau) < 0$ 
for a particle-antiparticle interaction, so that in (\ref{lor-coul-2}) we
will have $-e^{2}(1-g_{55}\frac{c_{5}}{c}) \rightarrow e^{2}(1+g_{55}\frac{c_{5}}{c})$.
This expression therefore leads to an experimental signature for the model, 
predicting a discrepancy between e$^-$/e$^-$ scattering and e$^+$/e$^-$ scattering at
extremely low energy, and provides an experimental bound on $c_5 / c$.

In order to understand the role of $c_5$ in electromagnetic interactions, we
study an arbitrary event $X^{\mu }\left( \tau \right) $, which induces the current
\begin{equation}
j_{\varphi }^{\alpha }\left( x,\tau \right) = c \int ds~\varphi \left( \tau-s\right) 
\dot{X}^{\alpha }\left( s\right) \delta ^{4}\left[ x-X\left(
s\right) \right] \ \ .
\end{equation}
The form of $G_{Maxwell}$ allows us apply standard techniques associated with the
Li\'{e}nard-Wiechert potential.  Writing
\begin{eqnarray}
a^{\alpha }\left( x,\tau \right) &&\mbox{\hspace{-20pt}}=-\frac{e}{c} \int
d^{4}x^{\prime }d\tau ^{\prime } \; G_{Maxwell}\left( x-x^{\prime } , \tau -\tau
^{\prime }\right) j_{\varphi }^{\alpha }\left( x^{\prime },\tau^{\prime }
\right) 
\notag \\
&&\mbox{\hspace{-20pt}}=\frac{e}{2\pi }\int ds~\varphi \left( \tau
-s\right) \dot{X}^{\alpha }\left( s\right) \delta \left( \left( x-X\left(
s\right) \right) ^{2}\right) \theta ^{ret}
\end{eqnarray}
and using the identity
\begin{equation}
\dint d\tau f\left( \tau \right) \delta \left[g\left( \tau \right) \right]
=\dfrac{f\left( \tau_R\right) }{\left\vert g^{\prime }\left( \tau_R\right)
\right\vert } \ \ ,
\label{identity}
\end{equation}
where $\tau_R$ is the retarded time found from
\begin{equation}
g\left( \tau \right) =(x-X(\tau_R))^2 =0 \qquad \qquad
\theta ^{ret}=\theta \left( x^{0}-X^{0}\left( \tau_R\right) \right) \ \ ,
\end{equation}
provides
\begin{equation}
a^{\alpha }\left( x,\tau \right) =\frac{e}{4\pi }\varphi \left( \tau -\tau_R\right) \frac{
\dot{X}^{\alpha }\left( \tau_R\right) }{\left\vert \left( x^{\mu }-X^{\mu }\left( \tau_R\right)
\right) \dot{X}_{\mu }\left( \tau_R\right) \right\vert} \ \ .
\end{equation}
Using this potential,
where we write the event velocity and line of observation as
\begin{equation}
u^\mu = \dot{X}^\mu (\tau) \qquad \qquad z^\mu = x^\mu - X^\mu(\tau)
\end{equation}
the potential takes the form
\begin{equation}
a^{\mu }\left( x,\tau \right) =\frac{e}{4\pi }\varphi \left( \tau -\tau_R\right) \frac{
u^{\mu } }{\left\vert u \cdot z \right\vert} \qquad \qquad 
a^{5 }\left( x,\tau \right) =\frac{e}{4\pi }\varphi \left( \tau -\tau_R\right) \frac{
c_5 }{\left\vert u \cdot z \right\vert}
\ \ .
\label{LV}
\end{equation}
%
%
By a similar procedure we find the field strengths, separated into the retarded and radiation parts, as
\begin{eqnarray}
f_{ret}^{\mu \nu }(x,\tau) \eq -\frac{e}{4\pi }\varphi \left( \tau -\tau_R\right) \frac{\left( z^{\mu
}u^{\nu }-z^{\nu }u^{\mu }\right) u^{2}}{\left( u\cdot z\right) ^{3}}
\sim \dfrac{1}{\mathbf{z}^{2}} \strt{18}
\label{rad-1}
\\
f_{ret}^{5\mu }(x,\tau)  \eq \frac{ec_5}{4\pi }\varphi \left( \tau
-\tau_R\right) \frac{z^{\mu }u^{2}-u^{\mu }\left( u\cdot z\right) }{\left(
u\cdot z\right) ^{3}}
\sim \dfrac{1}{\mathbf{z}^{2}}\strt{30}
\label{rad-2}
\\
f_{rad}^{\mu \nu }(x,\tau) \eq -\frac{e}{4\pi }\varphi (\tau -\tau_R)
\left[
\frac{\left( z^{\mu }\dot{u}^{\nu
}-z^{\nu }\dot{u}^{\mu }\right) \left( u\cdot z\right) -\left( z^{\mu
}u^{\nu }-z^{\nu }u^{\mu }\right) \left( \dot{u}\cdot z\right) }{
\left( u\cdot z\right) ^{3}} \right. \notag
\\
&&\qquad\qquad\qquad   +\left. \frac{\epsilon \left( \tau  -\tau_R\right) }{ \lambda } \frac{z^{\mu }u^{\nu
}-z^{\nu }u^{\mu }}{\left( u\cdot z\right) ^{2}} \right]
\sim \dfrac{1}{\left\vert 
\mathbf{z}\right\vert }\strt{24} 
\label{rad-3}
\\
f_{rad}^{5\mu }(x,\tau)  \eq -\frac{ec_5}{4\pi }\varphi \left( \tau
-\tau_R\right) \left[ \frac{\left( \dot{u}\cdot z\right) z^{\mu }}{\left( u\cdot
z\right) ^{3}} -\frac{\epsilon \left( \tau  -\tau_R\right) }{ \lambda }
\frac{z^{\mu }-u^{\mu }\left( u\cdot z\right) }{\left( u\cdot
z\right) ^{2}} \right]
\sim \dfrac{1}{\left\vert \mathbf{z}\right\vert } \ \ .
\label{rad-4}
\end{eqnarray}
We have used $\left\vert u\cdot z\right\vert = - \left(
u\cdot z\right)$ (easily seen in a co-moving frame) and
\begin{equation}
\frac{d}{d\tau_R}\varphi \left( \tau -\tau_R\right) 
= -\frac{1}{2 \alpha \lambda }\frac{d}{d\tau }
e^{-\left\vert \tau -\tau_R\right\vert / \alpha \lambda }=-\frac{\epsilon \left(
\tau  -\tau_R\right) }{ \alpha \lambda }\varphi
\left( \tau  -\tau_R\right)
\end{equation}
where $\epsilon\left( \tau \right) = \text{signum}(\tau)$.
Notice that the $\tau$-dependence in these expressions is limited to the smoothing function $\varphi \left( \tau -\tau_R\right)$
and again $\lambda$ plays the role of a correlation length that localizes the
interaction to the neighborhood $\tau_R \pm \lambda $.  As expected,
concatenation of the potentials and field strengths recovers the expressions
found in standard Maxwell theory.

In (\ref{rad-1}) to (\ref{rad-4}) we see once again 
that $c_5$ multiplies $f^{5\mu}$ and so provides a relative scale factor with
respect to the
components $f^{\mu\nu}$.  Using (\ref{lorentz}), (\ref{alpha}) and (\ref{inv}), the Lorentz force on an event
moving in the field induced by another event can be written
\begin{eqnarray}
M\ddot{x}^{\mu } \eq \dfrac{e_0}{c} \left[ f^\mu_{\; \;  \nu} (x,\tau )\dot{x}^\nu
+ f^{5\mu} (x,\tau )\dot{x}^5 \right]
\strt{18} 
\notag \\ 
\eq \dfrac{e_0}{c} \frac{e}{4\pi }\varphi \left( \tau -\tau_R\right) \left[
\mathcal{F}^\mu_{\; \;  \nu} (x,\tau )\dot{x}^\nu
+ c_5^2 \ \mathcal{F}^{5\mu} (x,\tau ) \right]
\strt{18} 
\notag \\
\eq \dfrac{e_0}{c} \frac{e}{4\pi }\frac{1}{2 \alpha \lambda }
e^{-\left\vert \tau -\tau_R\right\vert / \alpha \lambda } \left[
\mathcal{F}^\mu_{\; \;  \nu} (x,\tau )\dot{x}^\nu
+ c_5^2 \ \mathcal{F}^{5\mu} (x,\tau ) \right]
\strt{18} 
\notag \\
\eq \frac{e^2}{4\pi c} \ 
e^{-\left\vert \tau -\tau_R\right\vert / \alpha \lambda } \ \frac{1}{2\alpha }  \left[
\mathcal{F}^\mu_{\; \;  \nu} (x,\tau )\dot{x}^\nu
+ c_5^2 \ \mathcal{F}^{5\mu} (x,\tau ) \right]
\strt{18} 
\notag \\
\eq \frac{e^2}{4\pi c} \ 
e^{-\left\vert \tau -\tau_R\right\vert / \alpha \lambda } \ \frac{\mathcal{F}^\mu_{\; \;  \nu} (x,\tau )\dot{x}^\nu
+ c_5^2 \ \mathcal{F}^{5\mu} (x,\tau )}{1 + \left( 
c_5 / c \right)^2  }
\label{lorentz-2}
\end{eqnarray}
where
\begin{eqnarray}
\mathcal{F}^{\mu \nu }(x,\tau) \eq f^{\mu \nu }(x,\tau)
\strt{14} 
%
\label{rad-5}
\\
\mathcal{F}^{5\mu }(x,\tau)  \eq \frac{z^{\mu }u^{2}-u^{\mu }\left( u\cdot z\right) }{\left(
u\cdot z\right) ^{3}}-  \frac{\left( \dot{u}\cdot z\right) z^{\mu }}{\left( u\cdot
z\right) ^{3}}  + \frac{\epsilon \left( \tau  -\tau_R\right) }{ \lambda }
\frac{z^{\mu }-u^{\mu }\left( u\cdot z\right) }{\left( u\cdot
z\right) ^{2}} 
\strt{12}
\label{rad-6}
%
%
%
\end{eqnarray}
are independent of $c_5$.
The Lorentz force interaction will be in the range 
\begin{equation}
M\ddot{x}^{\mu }  = \frac{e^2}{4\pi c} \times
\left\{
\begin{array}{lrl}
e^{-2\left\vert \tau -\tau_R\right\vert /  \lambda } \ 
\mathcal{F}^\mu_{\; \;  \nu} (x,\tau )\dot{x}^\nu & ,
& c_5 \longrightarrow 0 \cr
& & \cr
e^{-\left\vert \tau -\tau_R\right\vert / \lambda } \ \ \frac{1}{2} \left[
\mathcal{F}^\mu_{\; \;  \nu} (x,\tau )\dot{x}^\nu
+  \mathcal{F}^{5\mu} (x,\tau ) \right]& ,
& c_5 \longrightarrow c
\end{array}
\right.
\end{equation}
showing that $c_5 / c$ provides a
continuous scaling of the Lorentz force.  Taking $c_5 \rightarrow 0$ reproduces
standard Maxwell dynamics in much the way that taking $c \rightarrow \infty$
reproduces nonrelativistic mechanics.  Unlike the nonrelativistic approximation,
in which the speed of light is taken to be infinite and action at a distance
instantaneous, in the Maxwell approximation of pre-Maxwell theory the event dynamics
evolve so slowly over $\tau$ that the system is essentially in equilibrium,
the event density vanishes and in particular, no mass exchange takes place.
This equilibrium is the spacetime generalization of a nonrelativistic static system.
The contribution associated with $G_{Correlation}$ is less straightforward,
but in Appendix B we gain some insight by studying the $\delta$-function term and assuming
that the structure of the $\theta$-function term must be sufficiently similar to
permit cancelation of singularities.  

\section{A self-interaction}

It was seen in (\ref{mass-shift}) that particles may exchange mass with the
fifth electromagnetic field through
\begin{equation*}
\frac{d}{d\tau }(-\tfrac{1}{2}M\dot{x}^{2})=  \dfrac{e_0}{c}\; f^{5\mu }\dot{x }_{\mu}
\end{equation*}
and despite the scaling of $f^{5\mu }$ by $c_5 / c$ this effect cannot be
assumed to be insignificant.  In order to account for the observed stability of
particle masses, we must find some mechanism that tends to enforce on-shell
evolution, perhaps by damping off-shell behavior in the manner of air friction
producing a terminal velocity.  If, for example, some circumstance were to
produce a field of the form $f^{5\mu } = \sigma \dot{x }^{\mu}$ then 
\begin{equation}
\frac{d}{d\tau }(-\tfrac{1}{2}M\dot{x}^{2})=  \dfrac{e_0}{c}\; \sigma
\dot{x }^{\mu}\dot{x }_{\mu} =  -\dfrac{2  e_0\sigma }{Mc} \left( - \tfrac{1}{2}M\dot{x
}^2 \right)
\end{equation}
producing mass decay.

In this section, we propose a model for a self-interaction between a moving
event and its electromagnetic field that produces a mass decay but vanishes for
on-shell propagation.  Unlike the self-interaction between a particle and its
radiation field, associated with the Abraham-Lorentz-Dirac equation, this model
involves the influence of the field induced through $G_{Correlation}$ with
retarded timelike support.  The event experiences a force along its worldline
produced by its earlier motion along that worldline.

\subsection{Framework}

As in Appendix B, we study the motion of an arbitrarily moving event $X^\mu
(\tau)$, this time in a co-moving frame, so that
\begin{equation}
X\left( \tau \right) =\left( ct\left( \tau \right) ,\mathbf{0}\right) %
\mbox{\qquad}\dot{X}\left( \tau \right) =\left( c\dot{t}\left( \tau \right) ,%
\mathbf{0}\right) \ \ .
\end{equation}
In this frame
\begin{equation}
\dot{X}^2 = -c^2 \dot{t}^2
\end{equation}
and so off-shell propagation is characterized by $\dot{t} \ne 1$ in the rest
frame. 
We are interested in the self-force on the event at time $\tau ^{\ast }$ and
write the observation point as
\begin{equation}
X(\tau^\ast) = (c t(\tau^\ast) , \mathbf{x}(\tau^\ast) )
\end{equation}
so that
\begin{equation}
X(\tau^\ast) - X(s) = (c t(\tau^\ast) , \mathbf{x}(\tau^\ast) ) - ( c t(s) ,
\mathbf{x}(s) ) = c( t(\tau^\ast) - t(s) , \mathbf{0}) \ \ .
\end{equation}
Because $G_{Maxwell} = 0$ on this timelike separation, the sole contribution
comes from $G_{Correlation}$.  As in Appendix A, we approximate 
%
%
$\varphi (\tau ^{\prime }-s)=\delta (\tau ^{\prime }-s)$,
%
%
introduce the function $g(s)$ to express terms of the type
\begin{equation}
c^2 g\left(s\right) = - \left( \left( X(\tau) -X(s) \right) ^{2}+c_{5}^{2}(\tau -s)^{2}\right) = 
c^2 \left( \left( t\left( \tau ^{\ast }\right) -t\left( s\right) \right)
^{2}-\frac{c_{5}^{2}}{c^{2}}(\tau ^{\ast }-s)^{2}\right) \ \ ,
\end{equation}
and write
\begin{equation}
a^{\alpha }\left( X\left( \tau ^{\ast }\right) ,\tau ^{\ast }\right) ={\frac{
ec_5}{{2\pi ^{2}}c^{3}}}\int ds\ \dot{X}^{\alpha }(s)\left( \frac{1}{2}{\frac{{
\theta }\left( g(s) \right) }{\left( g(s) \right) ^{3/2}}}-{\frac{{\delta }
\left( g(s) \right) }{\left( g(s) \right) ^{1/2}}}\right) \ \theta ^{ret}
\label{si_pot}
\end{equation}
for the self-field experienced by the event.  We designate the two terms as
\begin{equation}
a^{\alpha }\left( X\left( \tau ^{\ast }\right) ,\tau ^{\ast }\right) =
a^{\alpha }_\theta + a^{\alpha }_\delta
\end{equation}
\subsection{Uniform on-shell motion}

For an event evolving uniformly on-shell we have
\begin{equation}
t\left( \tau ^{\ast }\right) =\tau ^{\ast } \qquad \qquad g(s) = \left( 1
-\frac{c_{5}^{2}}{c^{2}}\right)(\tau ^{\ast }-s)^{2}
\end{equation}
and using identity (\ref{identity}) are led to
\begin{eqnarray}
a\left( X\left( \tau ^{\ast }\right) ,\tau ^{\ast }\right) \eq
\frac{ec_5}{2\pi^2c^3}
\left( c,\mathbf{0},c_{5}\right) \int ds \ \theta \left( \tau ^{\ast }-s\right)
\notag \\
&& \left( \frac{1}{2}
\frac{\theta\left(\left(1-\frac{c_5^2}{c^2}\right) \left(\tau^\ast-s\right)^2\right) }
{\left( \left( 1-\frac{c_5^2}{c^2}\right)\left(\tau ^\ast -s\right) ^2 \right) ^{3/2}}
-\frac{\delta \left(  \left( 1-\frac{
c_{5}^{2}}{c^{2}}\right)\left( {\tau ^{\ast }-s}\right) {^{2}}  \right) }{\left( \left( 1-\frac{
c_{5}^{2}}{c^{2}}\right)\left( {\tau ^{\ast }-s}\right) {^{2}} \right)
^{1/2}}\right) \strt{36} 
\notag \\
\eq {\frac{
ec_5\left( c,\mathbf{0},c_{5}\right)}{{2\pi ^{2}}c^{3}\left( 1-{\frac{c_{5}^{2}}{c^{2}}}%
\right) ^{3/2}
}} \int_{-\infty }^{{\tau ^{\ast }}}ds\ \left( \frac{1}{2}{%
\frac{{1}}{\left( {\tau ^{\ast }-s}\right) {^{3}}}}-{\frac{{\delta }\left( {%
\tau ^{\ast }-s}\right) 
\theta \left( {\tau ^{\ast }-s}\right)}{\left\vert \left( {\tau ^{\ast
}-s}\right) ^{2}\right\vert }}\right) \ \ .
\end{eqnarray}
Since
\begin{equation}
\int_{-\infty }^{{\tau ^{\ast }}}ds\ \frac{{1}}{\left( {\tau ^{\ast }-s}
\right) {^{3}}}=\left. {\frac{{1}}{2\left( {\tau ^{\ast }-s}\right) {^{2}}}}
\right\vert _{-\infty }^{{\tau ^{\ast }}}=
\lim_{s \rightarrow \tau^\ast}  \ \frac{{1}}{2\left( {\tau ^{\ast }-s}\right) {^{2}}}
\end{equation}
and
\begin{equation}
\int_{-\infty }^{{\tau ^{\ast }}}ds\ {\frac{{\delta }\left( {\tau ^{\ast }-s}
\right) \theta \left( {\tau ^{\ast }-s}\right) }{\left( {\tau ^{\ast }-s}
\right) ^{2}}
= \lim_{s \rightarrow \tau^\ast}  \ {\frac{\theta \left( {\tau ^{\ast }-s}\right) }{\left( {
\tau ^{\ast }-s}\right) ^{2}}}
= \lim_{s \rightarrow \tau^\ast}  \ \frac{ \frac{1}{2}}{\left( {\tau ^{\ast }-s}\right) ^{2}}}
\end{equation}
we find that for uniform on-shell motion
\begin{equation}
a\left( X\left( \tau ^{\ast }\right) ,\tau ^{\ast }\right) = \frac{ec_5}{2\pi^2c^3}
\left( c,\mathbf{0},c_{5}\right) 
\lim_{s \rightarrow \tau^\ast}  \ \left( 
\frac{{1}}{2\left( {\tau ^{\ast }-s} \right) {^{2}}} -\frac{ \frac{1}{2}}{\left( {\tau ^{\ast }-s}\right) ^{2}}
\right) = 0 \ \ .
\end{equation}
\subsection{Field strengths}

From $\dot X^i = 0$ and the form of (\ref{si_pot})
\begin{equation}
a^i = 0 \qquad \qquad \partial_i a^0 = \partial_i a^5 = 0 \qquad 
\Rightarrow \qquad f^{\mu\nu} = f^{5i} = 0
\end{equation}
and so the field reduces to
\begin{equation}
f^{50}=\partial ^{5}a^{0}-\partial ^{0}a^{5}=g^{55}\frac{1}{c_{5}}\partial
_{\tau ^{\ast }}a^{0}-g^{00}\frac{1}{c}\partial _{t}a^{5}=\frac{1}{c_{5}}
\partial _{\tau ^{\ast }}a^{0}+\frac{1}{c}\partial _{t}a^{5}
\end{equation}
where the partial derivative $\partial _{\tau ^{\ast }}$ only acts on the explicit
variable (not on $t\left( \tau ^{\ast }\right) $ or $\theta ^{ret}$).
Similarly, the velocity appears as $\dot X^\alpha (s)$ and is constant with
respect to $\partial _{\tau ^{\ast }}$. 

Working piece-by-piece
\begin{equation}
\partial ^{5}a_{\theta }^{0}=\frac{ec_{5}}{{4\pi ^{2}}c^{3}}\frac{1}{c_{5}}
\partial _{\tau ^{\ast }}\int ds\  c\dot{t}(s){\frac{{\theta }\left( \left( t\left( \tau
^{\ast }\right) -t\left( s\right) \right) ^{2}-\frac{c_{5}^{2}}{c^{2}}(\tau
^{\ast }-s)^{2}\right)}{\left[ \left( t\left( \tau ^{\ast
}\right) -t\left( s\right) \right) ^{2}-\frac{c_{5}^{2}}{c^{2}}(\tau ^{\ast
}-s)^{2}\right] ^{3/2}}}\theta ^{ret}
\end{equation}
contains
\begin{equation}
\partial _{\tau ^{\ast }}{\theta }\left( \left( t\left( \tau ^{\ast }\right)
-t\left( s\right) \right) ^{2}-\frac{c_{5}^{2}}{c^{2}}(\tau ^{\ast
}-s)^{2}\right) =-2\frac{c_{5}^{2}}{c^{2}}{\delta }\left( \left( t\left(
\tau ^{\ast }\right) -t\left( s\right) \right) ^{2}-\frac{c_{5}^{2}}{c^{2}}
(\tau ^{\ast }-s)^{2}\right) (\tau ^{\ast }-s)
\end{equation}
and
\begin{equation}
\partial _{\tau ^{\ast }}\frac{1}{\left[ \left( t\left( \tau ^{\ast }\right)
-t\left( s\right) \right) ^{2}-\frac{c_{5}^{2}}{c^{2}}(\tau ^{\ast }-s)^{2}%
\right] ^{3/2}} = 3\frac{c_{5}^{2}}{c^{2}}\frac{\tau ^{\ast }-s}{\left[ \left( t\left( \tau
^{\ast }\right) -t\left( s\right) \right) ^{2}-\frac{c_{5}^{2}}{c^{2}}(\tau
^{\ast }-s)^{2}\right] ^{5/2}} \ \ .
\end{equation}
Similarly,
\begin{equation}
\frac{1}{c}\partial _{t}a_{\theta }^{5}={\frac{
ec_{5}}{{4\pi ^{2}}c^{3}}}\frac{1}{c}\partial _{t\left( \tau ^{\ast }\right)
}\int ds\ c_{5}{\frac{{\theta }\left( {\left( t\left( \tau ^{\ast }\right)
-t\left( s\right) \right) ^{2}-\frac{c_{5}^{2}}{c^{2}}(\tau ^{\ast }-s)^{2}}
\right) }{\left( \left( t\left( \tau ^{\ast }\right) -t\left( s\right)
\right) ^{2}-\frac{c_{5}^{2}}{c^{2}}(\tau ^{\ast }-s)^{2}\right) ^{3/2}}}\
\theta ^{ret}
\end{equation}
contains
\begin{eqnarray}
\partial _{t\left( \tau ^{\ast }\right) }{\theta }\left( \left( t\left( \tau
^{\ast }\right) -t\left( s\right) \right) ^{2}-\frac{c_{5}^{2}}{c^{2}}(\tau
^{\ast }-s)^{2}\right) \eq 2\left( t\left( \tau ^{\ast }\right) -t\left(
s\right) \right) \times 
\notag \\ 
&&\hspace{-40pt} {\delta }\left( \left( t\left( \tau ^{\ast }\right)
-t\left( s\right) \right) ^{2}-\frac{c_{5}^{2}}{c^{2}}(\tau ^{\ast
}-s)^{2}\right)
\end{eqnarray}
\begin{equation}
\partial _{t\left( \tau ^{\ast }\right) }\frac{1}{\left[ \left( t\left( \tau
^{\ast }\right) -t\left( s\right) \right) ^{2}-\frac{c_{5}^{2}}{c^{2}}(\tau
^{\ast }-\tau ^{\prime })^{2}\right] ^{3/2}} =-3\frac{t\left( \tau ^{\ast }\right) -t\left( s\right) }{\left[ \left(
t\left( \tau ^{\ast }\right) -t\left( s\right) \right) ^{2}-\frac{c_{5}^{2}}{
c^{2}}(\tau ^{\ast }-\tau ^{\prime })^{2}\right] ^{5/2}}
\end{equation}
and
\begin{equation}
\partial _{t\left( \tau ^{\ast }\right) }\theta ^{ret}=\partial _{t\left(
\tau ^{\ast }\right) }\theta \left( t\left( \tau ^{\ast }\right) -t\left(
s\right) \right) =\delta \left( t\left( \tau ^{\ast }\right) -t\left(
s\right) \right) = 0 
\end{equation}
where the last expression vanishes because $t\left( \tau ^{\ast }\right) =
t\left( s\right)$ makes the argument of the $\theta$-function
negative.  Putting the pieces together we find
\begin{eqnarray}
\partial^5 a^0_\theta - \partial^0 a^5_\theta
\eq \frac{3ec_5}{{4\pi ^{2}}c^{3}}\frac{c_{5}}{c}\int ds\ \frac{
{\theta }\left( \left( t\left( \tau ^{\ast }\right) -t\left( s\right)
\right) ^{2}-\frac{c_{5}^{2}}{c^{2}}(\tau ^{\ast }-s)^{2}\right) }{\left[
\left( t\left( \tau ^{\ast }\right) -t\left( s\right) \right) ^{2}-\frac{
c_{5}^{2}}{c^{2}}(\tau ^{\ast }-s)^{2}\right] ^{5/2}}
\ \theta ^{ret}
\ \Delta\left(\tau^\ast ,s\right)
\notag \\
&&\hspace{-19pt}-\frac{ec_5}{{2\pi ^{2}}c^{3}}\frac{c_{5}}{c}\int ds\ \frac{{\delta }\left(
\left( t\left( \tau ^{\ast }\right) -t\left( s\right) \right) ^{2}-\frac{
c_{5}^{2}}{c^{2}}(\tau ^{\ast }-s)^{2}\right) }{\left[ \left( t\left( \tau
^{\ast }\right) -t\left( s\right) \right) ^{2}-\frac{c_{5}^{2}}{c^{2}}(\tau
^{\ast }-s)^{2}\right] ^{3/2}}
\ \theta ^{ret}
\ \Delta\left(\tau^\ast ,s\right)
\end{eqnarray}
where
\begin{equation}
\Delta\left(\tau^\ast ,s\right) = \dot{t}(s)(\tau ^{\ast }-s)-\left( t\left( \tau ^{\ast }\right) -t\left( s\right) \right) \ \ . 
\end{equation}
Similarly, the derivatives of $a_\delta$ produce
\begin{eqnarray}
\partial^5 a^0_\delta - \partial^0 a^5_\delta \eq -\frac{ec_{5}}{{2\pi ^{2}}c^{3}}\frac{c_{5}}{c}\int ds\ {
\frac{{\delta }\left( \left( t\left( \tau ^{\ast }\right) -t\left( s\right)
\right) ^{2}-\frac{c_{5}^{2}}{c^{2}}(\tau ^{\ast }-s)^{2}\right) }{\left(
\left( t\left( \tau ^{\ast }\right) -t\left( s\right) \right) ^{2}-\frac{
c_{5}^{2}}{c^{2}}(\tau ^{\ast }-s)^{2}\right) ^{3/2}}} \ \theta ^{ret} \ \Delta\left(\tau^\ast ,s\right) 
\notag \\
&&\hspace{-19pt}-\frac{ec_{5}}{{2\pi ^{2}}c^{3}}\frac{c_{5}}{c}\int ds\ {\frac{2{\delta }
^{\prime }\left( \left( t\left( \tau ^{\ast }\right) -t\left( s\right)
\right) ^{2}-\frac{c_{5}^{2}}{c^{2}}(\tau ^{\ast }-s)^{2}\right) }{\left(
\left( t\left( \tau ^{\ast }\right) -t\left( s\right) \right) ^{2}-\frac{
c_{5}^{2}}{c^{2}}(\tau ^{\ast }-s)^{2}\right) ^{1/2}}} \ \theta ^{ret} \ \Delta\left(\tau^\ast ,s\right)
\end{eqnarray}
and combining terms we find
\begin{equation}
f^{50}=f_{\theta }^{50}+f_{\delta }^{50}+f_{\delta ^{\prime }}^{50}
\end{equation}
where
\begin{eqnarray}
f_{\theta }^{50} \eq
\frac{3e}{4\pi ^2}\frac{c_5^2}{c^4}
\int ds\ \frac{
{\theta }\left( \left( t\left( \tau ^{\ast }\right) -t\left( s\right)
\right) ^{2}-\frac{c_{5}^{2}}{c^{2}}(\tau ^{\ast }-s)^{2}\right) }{\left[
\left( t\left( \tau ^{\ast }\right) -t\left( s\right) \right) ^{2}-\frac{
c_{5}^{2}}{c^{2}}(\tau ^{\ast }-s)^{2}\right] ^{5/2}} \ \theta ^{ret} \ \Delta\left(\tau^\ast ,s\right) 
\label{f-theta}
\\
f_{\delta }^{50} \eq
-\frac{e}{\pi ^2}\frac{c_5^2}{c^4}
\int ds\ \frac{{
\delta }\left( \left( t\left( \tau ^{\ast }\right) -t\left( s\right) \right)
^{2}-\frac{c_{5}^{2}}{c^{2}}(\tau ^{\ast }-s)^{2}\right) }{\left[ \left(
t\left( \tau ^{\ast }\right) -t\left( s\right) \right) ^{2}-\frac{c_{5}^{2}}{
c^{2}}(\tau ^{\ast }-s)^{2}\right] ^{3/2}} \ \theta ^{ret} \ \Delta\left(\tau^\ast ,s\right) 
\label{f-delta}
\\
f_{\delta ^{\prime }}^{50} \eq
-\frac{e}{\pi ^2}\frac{c_5^2}{c^4}
\int
ds\ {\frac{{\delta }^{\prime }\left( \left( t\left( \tau ^{\ast }\right)
-t\left( s\right) \right) ^{2}-\frac{c_{5}^{2}}{c^{2}}(\tau ^{\ast
}-s)^{2}\right) }{\left( \left( t\left( \tau ^{\ast }\right) -t\left(
s\right) \right) ^{2}-\frac{c_{5}^{2}}{c^{2}}(\tau ^{\ast }-s)^{2}\right)
^{1/2}}} \ \theta ^{ret} \ \Delta\left(\tau^\ast ,s\right)
\label{f-delta-tag}
\end{eqnarray}

Notice that if the particle remains at constant velocity (in any uniform
frame), then
\begin{equation}
x^0\left( \tau \right) =u^0\tau \Rightarrow \Delta\left(\tau^\ast ,s\right)  =\frac{u^0}{c}(\tau ^{\ast
}-s)-\left( \frac{u^0}{c}\tau ^{\ast }-\frac{u^0}{c}s\right) =0
\end{equation}
and so $f^{50}$ vanishes.  For any smooth $t\left( \tau \right) $, 
\begin{eqnarray}
t\left( \tau ^{\ast }\right) -t\left( s\right) \eq t\left( s\right) +
\dot{t}(s)(\tau ^{\ast }-s)+\frac{1}{2}\ddot{t}(s)(\tau ^{\ast
}-s)^{2}+o\left( (\tau ^{\ast }-s)^{3}\right) -t\left( s\right) 
\notag \\
\eq \dot{t
}(s)(\tau ^{\ast }-s)+\frac{1}{2}\ddot{t}(s)(\tau ^{\ast }-s)^{2}+o\left( (\tau ^{\ast }-s)^{3}\right)
\end{eqnarray}
so the function
\begin{equation}
\Delta\left(\tau^\ast ,s\right) = \dot{t}(s)(\tau ^{\ast }-s)-\left( t\left( \tau ^{\ast }\right) -t\left( s\right) \right) =-\frac{1}{2}
\ddot{t}(s)(\tau ^{\ast }-s)^{2}+o\left( (\tau ^{\ast }-s)^{3}\right)
\end{equation}
is nonzero only when the time coordinate accelerates in the rest frame, equivalent to a shift in the
particle mass.  

\subsection{Mass jump}

As a first order example, we consider a small, sudden jump in mass at $\tau = 0$ characterized by
\begin{equation}
t\left( \tau \right) =\left\{ 
\begin{array}{lll}
\tau & , & \tau <0 \vspace{8pt} \\ 
\left( 1+\beta \right) \tau & , & \tau >0
\end{array}
\right.
\qquad \Rightarrow \qquad
\dot{t}\left( \tau \right) =\left\{ 
\begin{array}{lll}
1 & , & \tau <0 \vspace{8pt} \\ 
1+\beta & , & \tau >0
\end{array}
\right.
\end{equation}
and calculate the self-interaction.  Since $\theta^{ret}$ enforces $t(\tau^\ast) > t(s)$, it follows that
\begin{equation}
\tau^\ast < 0 \quad \Rightarrow \quad s < 0 \quad \Rightarrow \quad \dot t (\tau^\ast) = t(s) = 1
\quad \Rightarrow \quad \Delta(\tau^\ast , s) = 0 \ \ .
\end{equation}
Similarly, 
\begin{equation}
\tau^\ast >0 \ \ \text{and} \ \ s > 0 \quad \Rightarrow \quad \dot t (\tau^\ast) = t(s) = 1+\beta
\quad \Rightarrow \quad \Delta(\tau^\ast , s) = 0 \ \ .
\end{equation}
But when $\tau^\ast >0$ and $s < 0$,
\begin{equation}
\Delta(\tau^\ast , s) = \dot{t}(s)(\tau ^{\ast }-s)-\left( t\left( \tau ^{\ast }\right) -t\left(
s\right) \right) =\left( \tau ^{\ast }-s\right) -\left[ \left( 1+\beta \right) \left( \tau
^{\ast }\right) -s\right] =-\beta \tau ^{\ast }
\end{equation}
and $f^{50}$ can be found from the contributions (\ref{f-theta}) -- (\ref{f-delta-tag}).  Writing
\begin{equation}
g\left( s\right) = \left( t\left( \tau ^{\ast }\right) -t\left( s\right)
\right) ^{2}-\frac{c_{5}^{2}}{c^{2}}(\tau ^{\ast }-s)^{2}= \left( \left( 1+\beta \right)  \tau ^{\ast }
-s\right) ^{2}-\frac{c_{5}^{2}}{c^{2}}(\tau ^{\ast }-s)^{2} 
\end{equation}
and solving for $g(s^\ast) = 0$, we find
\begin{equation}
s^{\ast } =\left( 1+\frac{\beta }{1-\dfrac{c_{5}}{c}}\right) \tau ^{\ast }>\tau ^{\ast }
\end{equation}
so that $g(s) >0$ for $s < 0 < \tau^\ast$ and there will be no contribution from
(\ref{f-delta}) or (\ref{f-delta-tag}).  Thus,
\begin{eqnarray}
f^{50} \eq f_{\theta }^{50} = (-\beta \tau ^{\ast })\frac{3e}{4\pi ^2}\frac{c_5^2}{c^4} \int_{-\infty}^0 ds\ \frac{ 1
}{\left[ \left( t\left( \tau ^{\ast }\right) -t\left( s\right) \right) ^{2}-\dfrac{
c_{5}^{2}}{c^{2}}(\tau ^{\ast }-s)^{2}\right] ^{5/2}}   
\strt{36}
\notag
\\
\eq
(-\beta \tau ^{\ast })\frac{3e}{4\pi ^2}\frac{c_5^2}{c^4} \int_{-\infty }^{0}ds\ \frac{1}{\left[ \left(
\left( 1+\beta \right) \tau
^{\ast }-s\right) ^{2}-\dfrac{c_{5}^{2}}{c^{2}}(\tau ^{\ast }-s)^{2}\right]
^{5/2}} \ \ .
\end{eqnarray}
Shifting the integration variable as $x = \tau^\ast - s$ the integral becomes
\begin{equation}
\int_{-\infty }^{0}ds\ \frac{1}{\left[ \left(
\left( 1+\beta \right) \tau
^{\ast }-s\right) ^{2}-\dfrac{c_{5}^{2}}{c^{2}}(\tau ^{\ast }-s)^{2}\right]
^{5/2}} = - \int_{\infty }^{\tau ^{\ast }} \frac{dx}{\left( Cx^{2}+Bx+A\right)
^{5/2}}
\end{equation}
where
\begin{equation}
C = 1-\frac{c_{5}^{2}}{c^{2}} \mbox{\qquad}B=2u \tau
^{\ast }\mbox{\qquad}A=\left( \beta \tau ^{\ast }\right) ^{2} \ \ .
\end{equation}
This integral can be evaluated using the well-known form \cite{CRC}
\begin{equation}
\int \frac{dx}{\left( Cx^{2}+Bx+A\right) ^{5/2}} = \frac{2(2Cx+B)}{3q\sqrt{Cx^{2}+Bx+A}}\left(
\frac{1}{Cx^{2}+Bx+A} +  \frac{8C}{q} \right)
\end{equation}
where
\begin{equation}
q = 4AC - B^2
\end{equation}
leading to
\begin{eqnarray}
- \int_{\infty }^{\tau ^{\ast }} 
\dfrac{dx}{\left( Cx^{2}+Bx+A\right) ^{5/2}} \eq
-
\dfrac{1}{3\left( \beta \tau ^{\ast }\right) ^{4}} \times  \strt{28}\notag \\
&&\hspace{-60pt}
\left[ 2
\dfrac{c^{4}}{
c_{5}^{4}}\left( 1-
\dfrac{c_{5}^{2}}{c^{2}}\right) ^{3/2}\left( 1-
\dfrac{ \left( 1 - \dfrac{c_{5}^{2}}{c^{2}} \right)^{1/2}
\left( 1+
\dfrac{\beta }{\left( 1-
\dfrac{c_{5}^{2}}{c^{2}}\right) } \right) }{\left[ 1+
\dfrac{
2\beta }{1-
\dfrac{c_{5}^{2}}{c^{2}}}+
\dfrac{\beta ^{2}}{1-
\dfrac{c_{5}^{2}}{
c^{2}}}\right] ^{1/2}}\right) \right. \strt{108}
\notag \\
&&\hspace{-60pt}\qquad \left. +
\dfrac{c^{2}}{c_{5}^{2}}
\dfrac{\beta
^{2}\left( 1+
\dfrac{c_{5}^{2}}{c^{2}}
\dfrac{u }{1-
\dfrac{c_{5}^{2}}{c^{2}}}
\right) }{\left( 1-
\dfrac{c_{5}^{2}}{c^{2}}\right) ^{1/2}\left[ 1+
\dfrac{
2\beta }{1-
\dfrac{c_{5}^{2}}{c^{2}}}+
\dfrac{\beta ^{2}}{1-
\dfrac{c_{5}^{2}}{
c^{2}}}\right] ^{3/2}}\right] \strt{88}
\end{eqnarray}
and the field strength in the form
\begin{equation}
f^{50} = \frac{e}{4\pi ^2}\dfrac{1}{c_5^2\left( \beta \tau ^\ast \right) ^3} \ 
Q \left( \beta,\dfrac{c_5^2}{c^2} \right)
\end{equation}
where $Q \left( \beta,\frac{c_5^2}{c^2} \right)$ is the positive, dimensionless factor
\vspace{6pt}
\begin{eqnarray}
Q\left( \beta ,\dfrac{c_{5}^{2}}{c^{2}}\right)  \eq \left[
2\left( 1-\dfrac{c_{5}^{2}}{c^{2}}\right) ^{3/2}\left( 1-\dfrac{\left( 1-%
\dfrac{c_{5}^{2}}{c^{2}}\right) ^{1/2}\left( 1+\dfrac{\beta }{\left( 1-%
\dfrac{c_{5}^{2}}{c^{2}}\right) }\right) }{\left[ 1+\dfrac{2\beta }{1-\dfrac{%
c_{5}^{2}}{c^{2}}}+\dfrac{\beta ^{2}}{1-\dfrac{c_{5}^{2}}{c^{2}}}\right]
^{1/2}}\right) \right. \rule[-68pt]{0pt}{68pt}  \nonumber \\
&&\qquad \qquad \left. +\dfrac{\beta ^{2}\ \dfrac{c_{5}^{2}}{c^{2}}\left( 1+%
\dfrac{c_{5}^{2}}{c^{2}}\dfrac{\beta }{1-\dfrac{c_{5}^{2}}{c^{2}}}\right) }{%
\left( 1-\dfrac{c_{5}^{2}}{c^{2}}\right) ^{1/2}\left[ 1+\dfrac{2\beta }{1-%
\dfrac{c_{5}^{2}}{c^{2}}}+\dfrac{\beta ^{2}}{1-\dfrac{c_{5}^{2}}{c^{2}}}%
\right] ^{3/2}}\right] 
\end{eqnarray}
%
which is seen to be finite for $c_5 < c$
\begin{eqnarray}
Q \left( \beta,\dfrac{c_5^2}{c^2} \right)
\underset{c_5 \rightarrow 0 }{\xrightarrow{\hspace*{1.5cm}}}
2\left( 1-\dfrac{1+\beta }{\left[ 1+2\beta+\beta ^{2}\right] ^{1/2}}\right) = 0
 \ \ .
\end{eqnarray}
Since $f^{\mu\nu} = 0$, the Lorentz force induced by this field strength is then
\begin{equation}
M\ddot{x}^{\mu } =e_{0}f^{\mu \alpha }\dot{x}_{\alpha }=e_{0}f^{\mu 5}\dot{
x}_{5}=-e_{0}f^{5\mu }\dot{x}_{5}=-g_{55}e_{0}f^{5\mu }\dot{x}
^{5}=-e_{0}f^{5\mu }c_{5}
\end{equation}
and since $f^{5i} = 0$
\begin{eqnarray}
M\ddot{x}^{i} \eq 0 \\
M\ddot{x}^{0} \eq -c_{5}e_{0}f^{50}= \left\{
\begin{array}{ccc}
0&,& \tau^\ast < 0\vspace{8pt} \\
-\dfrac{\lambda e^2}{4\pi ^2}\dfrac{1}{c_5\left( \beta \tau ^\ast \right) ^3} \ Q \left(
\beta,\dfrac{c_5^2}{c^2} \right)&,& \tau^\ast > 0\\
\end{array}
\right.
\end{eqnarray}
in which the factor $\lambda$ is an artifact of the approximation $\varphi (\tau ^{\prime
}-s)=\delta (\tau ^{\prime }-s)$.
Under the influence of a negative Lorentz force, the $0$-coordinate will decelerate until the event
returns to on-shell propagation, so that 
the function $\Delta(\tau^\ast,s)$ and field strength $f^{50}$ again vanish. Similarly, 
\begin{equation}
\frac{d}{d\tau }\left( -\frac{1}{2}M\dot{x}^{2}\right) =e_{0}f^{5\mu }\dot{x}
_{\mu }=e_{0}f^{50}\dot{x}_{0}=-e_{0}cf^{50}\dot{t}=-{\frac{\lambda e^2}{{4\pi
^{2}}}}\dfrac{c}{c_5^2\left( \beta \tau ^\ast \right) ^3} \ 
Q \left( \beta,\dfrac{c_5^2}{c^2} \right) \dot t
\end{equation}
so that the mass will damp back to the on-shell value.
Notice also that if $\beta <0$ then $f^{50}$ changes sign
so that the self-interaction results in damping or anti-damping to restore
on-shell behavior.

We see that the Lorentz force is singular at $\tau^\ast = 0$, the moment at
which the velocity and mass
jump discontinuously.  We expect that the force will be smooth for a smooth mass increase.
Although this model is approximate, it seems to indicate that the self-interaction of the event with
the field generated by its mass shift will restore the event to on-shell propagation.  Additional
work is needed to provide a more complete solution.  

\section{A simple model for mass shift}

We consider an event propagating uniformly on-shell as
\begin{equation}
x\left( \tau \right) =u\tau =\left( u^{0},\mathbf{u}\right) \qquad \qquad \qquad 
u^{2}=-c^{2}
\end{equation}
until it passes through a dense region of charged particles that induce a small stochastic
perturbation $X\left( \tau \right) $ such that
\begin{equation}
x\left( \tau \right) =u\tau +X\left( \tau \right) \ \ .
\end{equation}
If the typical distance scale between force centers is $d$ then the perturbation will be roughly periodic with characteristic period
\begin{equation}
\frac{d}{\left\vert \mathbf{u}\right\vert }=\frac{\text{very short
distance}}{\text{moderate velocity}}=\text{very short time,}
\end{equation}
fundamental frequency 
\begin{equation}
\omega _{0}=2\pi \frac{\left\vert \mathbf{u}\right\vert }{d}=\text{very high
frequency,}
\end{equation}
and amplitude on the order of
\begin{equation}
\left\vert X^{\mu }\left( \tau \right) \right\vert \sim \alpha d
\end{equation}
for some macroscopic factor $\alpha < 1$.  We may expand the perturbation in a Fourier series
\begin{equation}
X\left( \tau \right) =\text{Re}\sum_{n}a_{n}~e^{in\omega _{0}\tau }
\end{equation}
and write the four-vector coefficients as
\begin{equation}
a_{n} = \alpha d s_{n} = \alpha d\left( s_{n}^{0},\mathbf{s}_{n}\right) =\alpha d\left(
cs_{n}^{t},\mathbf{s}_{n}\right) 
\end{equation}
where the $s_{n}$ represent a normalized Fourier series ($s_{0}^{\mu }\sim 1$).
The perturbed motion
\begin{equation}
X\left( \tau \right) =\alpha d~\text{Re}\sum_{n}s_{n}^{\mu }~e^{in\omega
_{0}\tau }
\end{equation}
is seen to be of scale $d$, but the perturbed velocity
\begin{eqnarray}
\dot{x}^{\mu }\left( \tau \right)   \eq u^{\mu }+\dot{X}^{\mu }\left( \tau
\right) 
\notag
\\
\eq u^{\mu }+\alpha d~\text{Re}\sum_{n}n\omega _{0}~s_{n}^{\mu
}~ie^{in\omega _{0}\tau }
\notag
\\
 \eq u^{\mu }+\alpha d~\text{Re}\sum_{n}n\left( 2\pi \frac{\left\vert \mathbf{u
}\right\vert }{d}\right) s_{n}^{\mu }~ie^{in\omega _{0}\tau }
\notag
\\
 \eq u^{\mu }+\alpha \left\vert \mathbf{u}\right\vert ~\text{Re}\sum_{n}2\pi
n~s_{n}^{\mu }~ie^{in\omega _{0}\tau }
\end{eqnarray}
is of macroscopic scale. The unperturbed, on-shell mass is
\begin{equation}
m=-\frac{M\dot{x}^{2}\left( \tau \right) }{c^{2}}=M
\end{equation}
and the perturbed mass is
\begin{eqnarray}
m  \eq -\frac{M\dot{x}^{2}\left( \tau \right) }{c^{2}}=-\frac{M}{c^{2}}\left(
u+\alpha \left\vert \mathbf{u}\right\vert ~\text{Re}\sum_{n}2\pi
n~s_{n}~ie^{in\omega _{0}\tau }\right) ^{2} 
\notag \\
 \eq -\frac{M}{c^{2}}\left( u^{2}+\left( \alpha \left\vert \mathbf{u}
\right\vert ~\text{Re}\sum_{n}2\pi n~s_{n}~ie^{in\omega _{0}\tau }\right)
^{2}+2\alpha \left\vert \mathbf{u}\right\vert ~\text{Re}\sum_{n}2\pi
n~\left( u\cdot s_{n}\right) ~ie^{in\omega _{0}\tau }\right)  
\notag \\
 \eq -\frac{M}{c^{2}}\left( -c^{2}+2\alpha \left\vert \mathbf{u}\right\vert ~
\text{Re}\sum_{n}2\pi n~\left( u\cdot s_{n}\right) ~ie^{in\omega _{0}\tau }
\right. \notag \\
&& \qquad \qquad \left.
-\left( \alpha \left\vert \mathbf{u}\right\vert \right) ^{2}\text{Re}
\sum_{n,m}\left( 2\pi \right) ^{2}nm~s_{n}\cdot s_{m}\ e^{i\left( n+m\right)
\omega _{0}\tau }\right) 
\notag \\
m \eq M\left( 1-\frac{2\alpha \left\vert \mathbf{u}\right\vert }{c^{2}}\text{Re}
\sum_{n}2\pi n~\left( u\cdot s_{n}\right) \ ie^{in\omega _{0}\tau } \right.  \notag \\
&& \qquad \left. +\frac{
\alpha ^{2}\mathbf{u}^{2}}{c^{2}}\text{Re}\sum_{n,m}\left( 2\pi \right)
^{2}nm~s_{n}\cdot s_{m}\ e^{i\left( n+m\right) \omega _{0}\tau }\right) 
\end{eqnarray}
Evaluating the typical coefficients in the rest frame of the unperturbed
motion
\begin{eqnarray}
\frac{2\alpha \left\vert \mathbf{u}\right\vert }{c^{2}}2\pi n~\left( u\cdot
s_{n}\right)   \eq \frac{4\pi \alpha \left\vert \mathbf{u}\right\vert n}{c^{2}}
~\left( c,\mathbf{0}\right) \cdot \left( cs_{n}^{t},\mathbf{s}_{n}\right)
=-4\pi \alpha \left\vert \mathbf{u}\right\vert ~ns_{n}^{t} \\
\frac{\alpha ^{2}\mathbf{u}^{2}}{c^{2}}\left( 2\pi \right) ^{2}nm~s_{n}\cdot
s_{m}  \eq 
\left( 2\pi \right) ^{2}\alpha ^{2}\mathbf{u}^{2}~nm\left(
s_{n}^{t}s_{m}^{t}-\frac{\mathbf{s}_{n}\cdot \mathbf{s}_{m}}{c^{2}}\right) 
\end{eqnarray}
and neglecting the $\alpha ^{2}$ term, we find
\begin{equation}
m\simeq M\left( 1+4\pi \alpha \left\vert \mathbf{u}\right\vert \text{Re}
\sum_{n}n~s_{n}^{t}\ ie^{in\omega _{0}\tau }\right) 
\end{equation}
which expresses a mass shift as
\begin{equation}
m\longrightarrow m\left( 1+\frac{\Delta m}{m}\right) \mbox{\qquad}\frac{
\Delta m}{m}=4\pi \alpha \left\vert \mathbf{u}\right\vert \text{Re}
\sum_{n}n~s_{n}^{t}\ ie^{in\omega _{0}\tau }\ \ .
\end{equation}
Larger mass shifts can be observed if $\alpha > 1$ and the second order term in $\alpha^2$ becomes
significant.

\section{Summary}

In Section 2, we obtained the SHP electromagnetic theory in a form that explicitly includes the constants $c$
and $c_5$ associated with the Einstein time $t$ and the invariant $\tau$, and considers
phenomenology in the case of $c_5 < c$.  We see that the field $f^{5\mu}$ that permits exchange of
mass between particles and fields generally appears in proportion to $c_5$, effectively scaling this
effect.  We also that at very low-energy the scale of particle-particle scattering is proportional to $1-g_{55}(c_5 / c)$
while particle-antiparticle scattering scales as $1+g_{55}(c_5 / c)$, providing an experimental
limit on $c_5$.  The implicit assumption that $c_5 = c$ would have precluded $g_{55} = +1$
because it would prohibit the static Coulomb force.  The possibility that $c_5 < c$ thus permits
either signature $g_{55} = \pm 1$.  

In Section 3, we consider a classical self-interaction in which an propagating event experiences a
force as it passes through the electromagnetic field induced by its earlier motion along its
worldline.  Such an interaction is prohibited by the lightlike support of $G_{Maxwell}$, the Maxwell part of the
SHP Green's function, and by the spacelike support of $G_{Correlation}$ for $g_{55} = -1$.  However,
for $g_{55} = +1$ the support of $G_{Correlation}$ is timelike and includes the particle's own
future worldline.  We found that for uniform motion this self-interaction vanishes --- it depends on
the time acceleration $\ddot x^0$ in the rest frame of the particle, associated with shifting mass.
The Lorentz force acting on a particle that undergoes a discrete jump in $\dot x^0$ in the rest
frame was found to be negative, and so acts on the particle motion to oppose the time acceleration
and restore the particle to on-shell propagation (for which the interaction again vanishes).  This
self-interaction would appear to provide an underlying mechanism for the asymptotic on-shell behavior found
by Aharonovich and Horwitz in numerical solutions \cite{jigal}, and perhaps an explanation of the observed mass
stability of the known particles.

In Section 4, we discuss a simple model in which a uniformly moving on-shell particle enters a
region of densely packed charges and experiences a mass shift induced by a very small
perturbation, which nevertheless contributes a very high frequency stochastic velocity.  

Considerable work is still required to work out the details of these simple models.

\section*{Appendix A --- Coulomb potential from $\mathbf{G_{Correlation}}$}

We are interested in an event moving as
\begin{equation}
X=\left( c\tau ,\mathbf{0}\right) \mbox{\qquad}u^{2}=-c^{2}
\end{equation}
where we approximate 
\begin{equation}
\varphi (\tau ^{\prime }-s)=\delta (\tau ^{\prime }-s)
\end{equation}
so that
\begin{equation}
a^{\alpha }\left( x,\tau \right) = -e\int ds\ G\left( x-X(s),\tau -s\right) \ \dot{X}^{\alpha }(s) 
=\frac{e}{2\pi ^{2}} \ \dot{X}^{\alpha }(s)\int ds\ G\left( x-X(s),\tau
-s\right) \ \ .
\end{equation}
We introduce the function $g(s)$ to express terms of the type
\begin{equation}
-\left( \left( x -X(s) \right) ^{2}+c_{5}^{2}(\tau -s)^{2}\right) =
{-\left( \left(\left( ct,\mathbf{x} \right) -\left( cs,\mathbf{0} \right) \right) ^{2}+c_{5}^{2}(\tau -s)^{2}\right) }
=c^2 g\left(s\right)
\end{equation}
where
\begin{equation}
{g}\left( s\right) = {\left( t-s\right) ^{2}{-}\frac{R^{2}}{c^{2}}-\frac{
c_{5}^{2}}{c^{2}}(\tau -s)^{2}} 
= Cs^{2}+Bs+A
\end{equation}
and
\begin{equation}
\zeta^{2}={\frac{c_{5}^{2}}{c^{2}}} \qquad 
C=\left( 1{-\zeta}^{2}\right) \qquad B=-2\left( t-{\zeta}^{2}\tau \right) 
\qquad A=t^{2}{{-}\frac{R^{2}}{c^{2}}-\zeta^{2}}\tau ^{2} 
\end{equation}
so that the potential can be written as
\begin{equation}
a\left( x,\tau \right) = {\frac{ec_5}{2\pi ^{2}c^{3}}}\left( c,\mathbf{0},
c_5\right) \int ds\ \left[ {\frac{1}{2}}\frac{{\theta }\left( {g}
\left( s\right) \right) }{g^{3/2}\left( s\right) }{-\frac{{\delta }\left( {g}
\left( s\right) \right) }{{g}^{1/2}\left( s\right) }}\right] \theta \left(
t-s\right) \ \ .
\end{equation}
The zeros of $g\left( s\right) $ are found to be
\begin{equation}
s_{\pm } = \frac{-B\pm \sqrt{B^{2}-4AC}}{2C}
=\frac{\left( t-{\zeta^{2}}\tau \right) \pm \sqrt{\dfrac{R^{2}}{c^{2}}
\left( 1-\zeta^{2}\right) +\zeta^{2}\left( t-\tau \right) ^{2}}}{\left( 1{-\zeta^{2}}
\right) }
\label{roots}
\end{equation}
and since we assume $\zeta^{2}<1$ there will be roots for any values of $t$ and
$R$.  In addition, the condition $\theta^{ret} = \theta ( t - s)$ requires $t > s$.  

If $t<s_{-}$ then
\begin{equation}
t <\frac{\left( t-{\zeta^{2}}\tau \right) -\sqrt{\frac{R^{2}}{c^{2}}\left(
1-\zeta^{2}\right) +\zeta^{2}\left( t-\tau \right) ^{2}}}{\left(
1{-\zeta^{2}}\right) } \qquad 
\Rightarrow \qquad 
-\zeta^{2}\left( t-\tau \right) ^{2} >\frac{R^{2}}{c^{2}}
\end{equation}
and so $t \ge s_{-}$ becomes a condition of integration for the $\theta$ term.  Similarly, if $t>s_{+}$ then
\begin{equation}
t >\frac{\left( t-{\zeta^{2}}\tau \right) +\sqrt{\frac{R^{2}}{c^{2}}\left(
1-\zeta^{2}\right) +\zeta^{2}\left( t-\tau \right) ^{2}}}{\left( 1{-\zeta^{2}}\right) }
\qquad \Rightarrow \qquad 
-\zeta^{2}\left( t-\tau \right) > \frac{R^{2}}{c^{2}}
\end{equation}
leading to the condition
\begin{equation}
s_{-} \le t \le s_{+}
\end{equation}
from which
\begin{equation}
a\left( x,\tau \right) ={\frac{ec_5}{{2\pi ^{2}}c^{3}}}\left( 1,\mathbf{0},
\frac{c_{5}}{c}\right) \left( {\frac{1}{2}}\int_{-\infty }^{s_{-}}ds\frac{{1}
}{g^{3/2}\left( s\right) }{-\int_{-\infty }^{\infty }ds\frac{{\delta }\left( 
{g}\left( s\right) \right) }{{g}^{1/2}\left( s\right) }}\theta \left(
t-s\right) \right) \ \ .
\end{equation}
Using the well-known form \cite{CRC}
\begin{equation}
\int \frac{dx }{(Cx^2 + Bx + A)^{3/2}} = \frac{
2\left( 2Cs+B\right) }{q(Cx^2 + Bx + A)^{1/2} }
\end{equation}
where
\begin{equation}
q = 4AC-B^2
\end{equation}
we notice from (\ref{roots}) that 
\begin{equation}
s_{-} =\frac{-B-\sqrt{B^{2}-4AC}}{2C}=\frac{-B-\sqrt{-q}}{2C} \qquad \Rightarrow \qquad 
-\sqrt{-q} = 2Cs_{-}+B
\end{equation}
and so
\begin{eqnarray}
\frac{1}{2}\int_{-\infty }^{s_{-}}ds\frac{{1}}{g^{3/2}\left( s\right) } \eq 
\frac{2Cs_{-}+B}{qg^{1/2}\left( s_{-}\right) }-\left. \frac{2Cs+B}{
qg^{1/2}\left( s\right) }\right\vert _{-\infty } \notag \\
\eq \frac{-\sqrt{-q}}{
qg^{1/2}\left( s_{-}\right) }+\frac{2\sqrt{C}}{\left(
2Cs_{-}+B\right) ^{2}} \notag \\
\eq \frac{1}{\sqrt{-q}g^{1/2}\left( s_{-}\right) }+\frac{1}{2}\frac{\sqrt{1{-\zeta}^{2}}}{\frac{R^{2}}{c^{2}}\left(
1-\zeta^{2}\right) +\zeta^{2}\left( t-\tau \right) ^{2}}
\end{eqnarray}
The second term is
\begin{equation}
\int_{-\infty }^{\infty }ds\frac{{\delta }\left( {g}\left( s\right) \right) 
}{{g}^{1/2}\left( s\right) }\theta \left( t-s\right)
\end{equation}
and using the identity
\begin{equation}
\int ds~f\left( s\right) ~\delta \left( g\left( s\right) \right) =\left. 
\frac{f\left( s^{-}\right) }{\left\vert g^{\prime }\left( s^{-
}\right) \right\vert }\right\vert _{s^{-}=g^{-1}\left( 0\right) }
\end{equation}
we can evaluate
\begin{equation}
{\int_{-\infty }^{\infty }ds\frac{{\delta }\left( {g}\left( s\right) \right) 
}{{g}^{1/2}\left( s\right) }}\theta \left( t-s\right) =\frac{\theta \left(
t-s_{-}\right) }{\left\vert {g}^{\prime }\left( s_{-}\right) \right\vert{g}^{1/2}\left( s_{-}\right) }
=\frac{1}{\left\vert { g}^{\prime }\left( s_{-}\right) \right\vert
{g}^{1/2}\left( s_{-}\right) } \ \ .
\end{equation}
Since
\begin{equation}
{g}^{\prime }\left( s_{-}\right) =\left( Cs_{-}^{2}+Bs_{-}+A\right) ^{\prime
}=2Cs_{-}+B=-\sqrt{-q}
\end{equation}
we see that this term cancels the singularity in the first term, leaving
\begin{equation}
{\frac{1}{2}}\int_{-\infty }^{s_{-}}ds\frac{{1}
}{g^{3/2}\left( s\right) }{-\int_{-\infty }^{\infty }ds\frac{{\delta }\left( 
{g}\left( s\right) \right) }{{g}^{1/2}\left( s\right) }}\theta \left(
t-s\right) = \frac{1}{2}\frac{\sqrt{1{-\zeta}^{2}}}{\frac{R^{2}}{c^{2}}\left(
1-\zeta^{2}\right) +\zeta^{2}\left( t-\tau \right) ^{2}}
\end{equation}
and
\begin{equation}
a\left( x,\tau \right)  =\frac{e}{{4\pi ^{2}}}\left( c,\mathbf{0},
c_5\right) \dfrac{c_5}{c}\dfrac{\sqrt{1-\dfrac{c_5}{c}}} 
{R^2\left( 1-\dfrac{c_5}{c}\right) +\dfrac{c_5}{c} c^2\left( t-\tau \right)
^{2}} \ \ .
\end{equation}
We notice that the potential has units of $c / $distance$^2 = 1 / $time $\times$
distance, as does the potential associated with $G_{Maxwell}$.  On
concatenation --- integration over $\tau$ --- we recover the 1/distance units of the
Maxwell potential.  This contribution to the potential is smaller by a factor of
$c_5 / c$ than the Yukawa potential found in (\ref{yukawa}), and drops off
faster with distance.

\section*{Appendix B --- $\mathbf{c_5}$-dependence of general potential from $\mathbf{G_{Correlation}}$}

We are interested in an arbitrary event moving as
\begin{equation}
X \left( \tau \right)  =\left( ct\left( \tau \right) ,\mathbf{x}\left( \tau
\right) \right)  \qquad \qquad X^5 = c_5 \tau
\end{equation}
and the induced field
\begin{equation}
a^{\alpha }\left( x,\tau \right) =-e\int ds\ G_{\varphi }\left( x-X(s),\tau
-s\right) \ \dot{X}^{\alpha }(s) \ \ .
\end{equation}
Making the approximation
\begin{equation}
\varphi (\tau ^{\prime }-s)=\delta (\tau ^{\prime }-s)
\end{equation}
leads to
\begin{eqnarray}
a^\alpha \left( x,\tau \right) \eq {\frac{ec_5}{{2\pi ^{2}}}}\int ds\ \dot{X}^\alpha \left(
s\right) \left( \frac{1}{2}\frac{{\theta }\left( {-\left( x-X\left(
s\right) \right) ^{2}-c_{5}^{2}(\tau -s)^{2}}\right) }{\left[ {-\left(
x-X\left( s\right) \right) ^{2}-c_{5}^{2}(\tau -s)^{2}}\right] ^{3/2}} \right.
\strt{24}
\notag \\
&& \qquad \qquad  \qquad \qquad \left.  -\frac{
{\delta }\left( {-\left( x-X\left( s\right) \right) ^{2}-c_{5}^{2}(\tau
-s)^{2}}\right) }{\left( {-\left( x-X\left( s\right) \right)
^{2}-c_{5}^{2}(\tau -s)^{2}}\right) ^{1/2}}\right) \theta ^{ret}
\end{eqnarray}
We designate
\begin{equation}
g\left( s\right)  ={-\left( x-X\left( s\right) \right) ^{2}-c_{5}^{2}(\tau
-s)^{2}} \qquad \qquad s^{\pm } =g^{-1}\left( 0\right) 
%
\end{equation}
and assume that the observation point $x$ is in the timelike future of
$X\left( -\infty \right) $ so that
\begin{eqnarray}
a^\alpha \left( x,\tau \right)  \eq \; {\frac{e}{{2\pi ^{2}}}}\int_{-\infty }^{s^{\pm
}}ds\ {\frac{1}{2}\dot{X}^\alpha \left( s\right) \frac{{\theta }\left( {-\left(
x-X\left( s\right) \right) ^{2}-c_{5}^{2}(\tau -s)^{2}}\right) }{\left[ {
-\left( x-X\left( s\right) \right) ^{2}-c_{5}^{2}(\tau -s)^{2}}\right] ^{3/2}
}}\theta \left( ct-X^{0}\left( s\right) \right)  
\notag \\
&&\hspace{-40pt} {-{\frac{e}{{2\pi ^{2}}}}\int_{-\infty }^{\infty }ds\ \dot{X}^\alpha \left(
s\right) \frac{{\delta }\left( {-\left( x-X\left( s\right) \right)
^{2}-c_{5}^{2}(\tau -s)^{2}}\right) }{\left( {-\left( x-X\left( s\right)
\right) ^{2}-c_{5}^{2}(\tau -s)^{2}}\right) ^{1/2}}}\theta \left(
ct-X^{0}\left( s\right) \right) \label{70}
\end{eqnarray}
Using the identity 
\begin{equation}
\int ds~f\left( s\right) \delta \left[ g\left( s\right) \right] 
=\sum_{s^\pm = g^{-1}\left( 0\right) }\frac{f\left( s\right) }{\left\vert
g^{\prime }\left( s\right) \right\vert }
\end{equation}
the second term in the integral becomes
\begin{equation}
%
{
\int_{-\infty }^{\infty }ds\ \dot{X}\left( s\right) \frac{{\delta }\left(
g\left( s\right) \right) }{\left( g\left( s\right) \right) ^{1/2}}}\theta
\left( ct-X^{0}\left( s\right) \right) =\frac{\dot{X}{\left( s^{\pm
}\right) }\theta \left( ct-X^{0}\left( s^{\pm }\right) \right) }{\left(
g\left( s^{\pm }\right) \right) ^{1/2}\left\vert g^{\prime }\left( s^{\pm
}\right) \right\vert } \label{72} \ \ .
\end{equation}
At the observation point $(x,x^5) = (x,c_5 \tau)$ we define a 5D line of observation as
\begin{equation}
Z = \left( z,z^5 \right) = (x,x^5) - \left( X\left( s\right), X^5 \right) =
\left( x - X(s) , c_5 \tau - c_5 s \right)
\end{equation}
and a 5-velocity
\begin{equation}
U = (u,u^5) \qquad \qquad u^\mu=\dot{X}^\mu\left( s\right) \qquad \qquad u^5 = \dot{X}^5
\end{equation}
%
leading to a generalization of the denominator of (\ref{LV}) in the form
\begin{equation}
g^{\prime }\left( s\right) = 2U\cdot Z
\end{equation}
so that (\ref{72}) becomes
\begin{equation}
%
%
\left. 
\frac{\dot{X}{\left( s\right) }\theta \left( ct-X^{0}\left( s\right)
\right) }{\left( g\left( s\right) \right) ^{1/2}\left\vert 2U\cdot
Z\right\vert }\right\vert _{s\longrightarrow s^{\pm }}
\end{equation}
%
which is singular as $s\rightarrow s^{\pm }$.  As seen in Appendix A, we expect that this singularity
is canceled by a corresponding singularity in the $\theta$-term of (\ref{70}).
Nevertheless, for $s\ne s^{\pm }$ this expression remains finite if we take ${c_{5}\rightarrow 0}$. 
Since we expect the $\delta $-term to have a
similar structure to the $\theta $-term, it seems that the contribution of $
G_{correlation}$ to the field induced by a general event will split as
\begin{equation}
\frac{\mathcal{F}_{\;\;\nu }^{\mu }(x,\tau )\dot{x}^{\nu }+c_{5}^{2}\ 
\mathcal{F}^{5\mu }(x,\tau )}{1+\left( c_{5}/c\right) ^{2}}
\end{equation}
where $\mathcal{F}_{\;\;\nu }^{\mu }(x,\tau )$ and $\mathcal{F}^{5\mu
}(x,\tau )$ remain finite as $c_{5} \rightarrow 0$.

\section*{Acknowledgments}

The author gratefully acknowledges the inspiration and guidance of Professor
Lawrence P. Horwitz, whose work has opened many fields of inquiry and whose
direction has set many students on a productive paths in physics.  The author
is also grateful to Mark Davidson for useful conversations on the nature mass
variation in SHP.


%
%
%

%
\end{document}